# Toward a Designable Extracellular Matrix: Molecular Dynamics Simulations of an Engineered Laminin-mimetic, Elastin-like Fusion Protein

James D. Tang[1], Charles E. McAnany[2], Cameron Mura[2], Kyle J. Lampe[1*]

[1]Department of Chemical Engineering, University of Virginia, Charlottesville, VA 22904, USA
[2]Department of Chemistry, University of Virginia, Charlottesville, VA 22904, USA

Keywords: elastin; laminin; extracellular matrix; brain; molecular simulation; lower critical solution temperature

## Abstract

Native extracellular matrices (ECMs) exhibit networks of molecular interactions between specific matrix proteins and other tissue components. Guided by these naturally self-assembling supramolecular systems, we have designed a matrix-derived protein chimera that contains a laminin globular–like (LG) domain fused to an elastin–like polypeptide (ELP). This bipartite design offers a flexible protein engineering platform: (i) laminin is a key multifunctional component of the ECM in human brains and other neural tissues, making it an ideal bioactive component of our fusion, and (ii) ELPs, known to be well-tolerated *in vivo*, provide a self-assembly scaffold with tunable physicochemical (viscoelastic, thermoresponsive) properties. Experimental characterization of novel proteins is resource-intensive, and examining many conceivable designs would be a formidable challenge in the laboratory. Computational approaches offer a way forward: molecular dynamics (MD) simulations can be used to analyze the structural/physical behavior of candidate LG-ELP fusion proteins, particularly in terms of conformational properties salient to our design goals, such as assembly propensity in a temperature range spanning the inverse temperature transition. As a first step in examining the physical characteristics of a model LG-ELP fusion protein, including its temperature-dependent structural behavior, we simulated the protein over a range of physiologically-relevant temperatures (290-320 K). We find that the ELP region, built upon the archetypal $(VPGXG)_5$ scaffold, is quite flexible, and has a propensity for $\beta$-rich secondary structures near physiological (310-315 K) temperatures. Our trajectories indicate that the temperature-dependent burial of hydrophobic patches in the ELP region, coupled to the local water structure dynamics and mediated by intramolecular contacts between aliphatic side-chains, correlates with the temperature-dependent structural transitions in known ELP polymers. Because of the link between compaction of ELP segments into $\beta$-rich structures and differential solvation properties of this region, we posit that future variation of ELP sequence and composition can be used to systematically alter the phase transition profiles, and thus the general functionality, of our LG-ELP fusion protein system.

## INTRODUCTION

A major challenge in neural tissue engineering and regenerative medicine is one of tissue construction: what biomaterial, in terms of chemical composition and physical properties, might best mimic the native extracellular matrix (ECM) that houses neural stem cells (NSCs), neurons, glia, and other cells? Engineered proteins afford an opportunity to systematically control both biological functionality and the structural/mechanical properties of the resulting ECM mimetic, thus enabling one to guide the behavior of encapsulated cells.[1, 2] For instance, neural cells encapsulated in engineered protein or peptide materials extend neurites hundreds of microns into the surrounding 3D matrix.[3] These materials permit cellular remodeling and bioresorption via cell-controlled proteolytic degradation, and inherently behave in a more physiologically native manner than do other biomimetics (e.g. commonly-used synthetic hydrogels). Tissue engineering can benefit immensely from artificial ECMs designed from naturally-occurring protein sequences: such polymers promote native cellular interactions and elicit desired regenerative behaviors *in vivo*[4, 5] while enabling control over bioactive and structural properties (porosity, proteolytic remodeling, cellular adhesion, stiffness, etc.). In short, biologically-based ECM mimetics provide a suitable matrix for the controlled organization of viable cells into physiologically-relevant tissues.[6, 7]

The ECM in neural tissue is a hierarchically structured composite material, consisting of proteoglycans and the large (typically >400 kDa) structural proteins collagen, fibronectin, and laminin. In the central nervous system (CNS), laminin is a particularly vital component of the ECM.[8, 9] Following a neural tissue injury, temporal regulation of laminin expression is critical in the production of potential neurotrophic and neurite-promoting factors by reactive astrocytes.[10]

Laminin also plays an important role in axonal growth in the developing mammalian CNS and in concurrent mechanotransduction events, such as in astrocyte cell adhesion and spreading.[8, 11]

Laminins are glycoproteins that provide a key linkage between cells and the broader ECM scaffold. Human laminin is an immense (900 kDa), disulfide-linked heterotrimer that consists of many globular domains and α–, $\beta$–, ϒ–rod-like chains; together, these entities assemble into a four-armed cruciform shape.[12] Several adhesion peptides have been identified within the laminin amino acid sequence; in particular, the [1124]RGD, [925]YIGSR and [2101]IKVAV segments are known recognition sites for as many as 20 integrins,[13] the 67 kDa laminin-1 receptor[14] and the 110-kDa laminin-binding protein,[15] respectively. These recognition sequences have been used to functionalize non-adhesive polymeric scaffolds, such as in hydrogels based on polyethylene glycol or hyaluronic acid.[16-18] However, these short ECM-derived peptide fragments are often imperfect in mediating cell-signaling events in neural tissue (cell attachment, axonal growth, etc.), likely because of (i) insufficient binding with cell-surface receptors and (ii) failure to initiate anchoring for assembly of basement membrane scaffolds.[19-23]

The fifth globular domain from the C-terminal region of the laminin α2 chain, denoted 'LG5', plays a key role as a binding site for integrins, heparin, and α-dystroglycan (α-DG).[24-27] Heparin is a highly anionic, polysulfated glycosaminoglycan (GAG) that binds exogenous growth factors and thereby helps regulate and maintain neural stem cell (NSC) differentiation.[28, 29] In neural cells, the α-DG glycoprotein complex plays a fundamental role in facilitating new laminin polymerization at the cell surface and in supporting cellular adhesion.[30, 31] LG5 also contains a region that binds integrin $\beta$1,[25, 27] which is part of an integrin adhesive complex that links the cytoskeleton and the ECM. Past work has focused on engineering hydrogels that contain only the short integrin-binding peptides from LG modules. A more effective biomimicry strategy might incorporate longer laminin sequences, enabling multifunctional biomaterials with native-like cell-binding capacities and targeted selectivity for growth factors (which, in turn, initiate stem cell self-renewal and differentiation programs). There is a precedent for engineering proteins functionalized with the LG5 domain to mediate cellular behavior.[32, 33] A yet further design criterion for ECM-mimetic fusion proteins is that they contain regions that enable assembly into higher-order structures, via either noncovalent (self-assembly) or covalent (chemical crosslinking) mechanisms. Elastin-like polypeptides (ELP) have generated much interest in the tissue engineering field, as the hierarchical self-assembly of these relatively ordered (via local interactions) peptides provides structural support in ECM materials, as well as the ability to control viscoelastic properties. The ability to tune the physical properties of ELP-containing regions offers a versatile way to modulate protein-mediated interactions between cells and the ECM—interactions that are critical in cellular adhesion, spreading, and migration.

ELPs undergo thermally-triggered first-order phase transitions,[34] characterized by a system-specific transition temperature known as the 'lower critical solution temperature' (LCST). This behavior is also termed an 'inverse temperature transition' as the polymer becomes *more* structured upon reaching the LCST, and separates into polymer-rich and water-rich phases. The solution behavior at/near the LCST depends on both (i) intrinsic factors, such as the amino acid composition[35, 36] and the number of $(VPGXG)_n$ pentapeptide repeats ('X' denotes a 'guest' residue, which can vary from one repeat to another), as well as (ii) extrinsic parameters, such as the concentration, pH, ionic strength, and other bulk solution properties.[37-42] Both sets of factors are useful in the context of protein design and engineering, as they are entirely manipulable: various ELP regions can be fused to a target protein and combined with systematic perturbation of experimental conditions to modulate protein/solution properties at and near the LCST. The assembly behavior at the LCST has been introduced into otherwise soluble polypeptides by fusing them to ELP regions.[43, 44] The thermoresponsive behavior of recombinant ELP fusions then allows simple purification via inverse transition cycling,[37] thus obviating expensive chromatographic resins and enabling large-scale production. Also, biocompatibility of ELP fusion proteins with biomechano-responsive properties has recently been demonstrated in animals.[45]

Fundamental progress in biomaterials discovery has been limited by a lack of high-resolution data about the structural dynamics of the underlying polymeric network. The properties of any material ultimately stem from the three-dimensional (3D) structures and dynamics of its molecular constituents—from the level of individual proteins to their higher–order assembly into matrices. These structural and dynamical properties, in turn, are deeply linked to the patterns of intra- and inter-molecular interactions that are thermodynamically accessible (and substantially populated) under a given set of experimental conditions. The structural and thermodynamic properties of a fusion protein design can be quantitatively characterized via experimental means (e.g., X-ray scattering), but systematically doing so on the scale of many dozens or even hundreds of designs would be prohibitively laborious and resource-intensive. Moreover, such approaches do not, in general, provide the atomic-resolution information on structure and dynamics that we need in order to iteratively refine and systematically improve our designs.

The thermodynamic properties and structural dynamics of various ELPs, above and below their LCSTs, have been studied by experimental and computational means.[46-50] However, a universally accepted, atomically-detailed description of the physicochemical and structural basis of this phase transition remains elusive;[48, 51, 52] also, past studies have generally examined short ELP regions in isolation, not fused to other protein domains. Deeper knowledge of the phase behavior and interfacial properties of ELPs would expand their general utility in biomaterial applications, and would mitigate the costs of producing and characterizing what end up being poorly structured (or otherwise undesirable) ECM candidates. Here, we have designed and simulated a multifunctional fusion construct, with the ultimate goal of driving neural differentiation via an engineered ECM that assembles under cyto-compatible conditions. We use the LG5 domain to supply crucial cell-protein-matrix interactions, while the ELP component of our modular design provides control over desired micro- and nanostructures. Being able to control the properties of our fusion goes in tandem with the architecture and physical properties of these matrices being stimuli-responsive, so environmental parameters such as temperature must be able to modulate the individual protein structures that compose such a matrix.

Using classical, all-atom MD simulations[53] we have examined the behavior of our LG-ELP design near its putative phase transition, as well as the temperature-dependent conformational and structural dynamics leading up to the LCST. These simulations supply picosecond-resolved, atomically-detailed information on discrete structural and functional states for our protein, on the overall timescale of ca. 100 ns. Thus, we can both analyze the molecular events near the presumed LCST transition of our fusion protein and also obtain an *a priori* view of the structural properties of our design, before dedicating experimental resources to the synthesis and characterization of a novel biopolymer with unknown (and otherwise unpredictable) LCST behavior.

## METHODS OF PROCEDURE

**LG-ELP fusion protein design methodology.** We designed an LG5–ELP fusion protein with the intention that it be able to undergo a temperature-induced structural transition, leading to formation of a functional ECM suitable for CNS tissue regeneration. Four design criteria were applied: (i) The fusion protein should be thermodynamically stable (i.e. retain native structure) under physiological conditions (temperature, pH, ionic strength). (ii) The fusion protein should feature bioactive sites along the LG portion of the peptide chain, and the ELP must not interact with the LG portion in a manner that occludes these bioactive sites (proteolytic sites, cell-binding domains, binding sites for other ECM molecules or growth factors, etc.). (iii) The fusion protein should be capable of self-assembly via noncovalent interactions. (iv) The self-assembly properties should be readily controllable by altering the assembly-driving sequence element (ELP, in our case).

ELPs consist of a pentapeptide repeat, $(VPGXG)_n$, where X is any guest residue other than proline. ELPs are described using the notation "$ELP[W_iY_jZ_k]_n$", where *W*, *Y*, and *Z* are the single-letter codes for the amino acids at X, the subscripts *i*, *j* and *k* indicate the number of pentamers featuring that guest residue, and *n* is the total number of repeats. From our estimates using the $T_t$-based hydrophobicity scale of amino acids,[50, 54, 55] and the LCST behavior of various other engineered ELP fusions,[44, 56] we designed an ELP with the sequence $ELP[K_2L_2I_2K_2]_1$. We predict that this motif will satisfy the aforementioned design criteria. The repeated Gly-Leu and Gly-Ile dipeptides serve as cleavage sites for type IV collagenase (gelatinase),[57] rendering the hydrogel susceptible to enzymatic cleavage and thereby allowing cell spreading and migration. In addition, the primary amine functionality of the lysine side-chain ($\varepsilon$-$NH_3^+$) enables site-specific coupling or crosslinking reactions.[58]

**MD simulations of LG-ELP.** Our LG-ELP design fuses the LG5 domain, known to adopt an antiparallel *β*-sandwich fold, to a C′-terminal ELP tail (Figure 1). Our starting 3D model for the LG5 domain was drawn from the crystal structure of the mouse homolog of the laminin α2 chain (PDB 1DYK),[19] which contains residues 2934–3117 of that particular laminin chain. An initial 3D structure for the 42-residue ELP-$[K_2L_2I_2K_2]_1$ sequence, $(\mathrm{GVG})(\mathrm{VPG}\underline{\mathrm{K}}\mathrm{G})_2(\mathrm{VPG}\underline{\mathrm{L}}\mathrm{G})_2(\mathrm{VPG}\underline{\mathrm{I}}\mathrm{G})_2(\mathrm{VPG}\underline{\mathrm{K}}\mathrm{G}\text{-}\mathrm{VPG}\underline{\mathrm{K}})$, was built using the peptide builder tool in the program Avogadro;[46] the N′-terminal GVG in the above sequence is a linker from the C′-terminus of the LG5 domain. The ELP starting structure was modeled as a canonical α-helix, with backbone torsion angles of φ = –60°, ψ= –40° (Figure S1). ELPs are likely only loosely structured in solution,[59] so the helical starting structure was not anticipated to bias the equilibrium structural ensemble (at least not if given sufficient sampling). Atomistic MD simulations were performed in NAMD, under the CHARMM36 force-field for proteins.[60, 61]

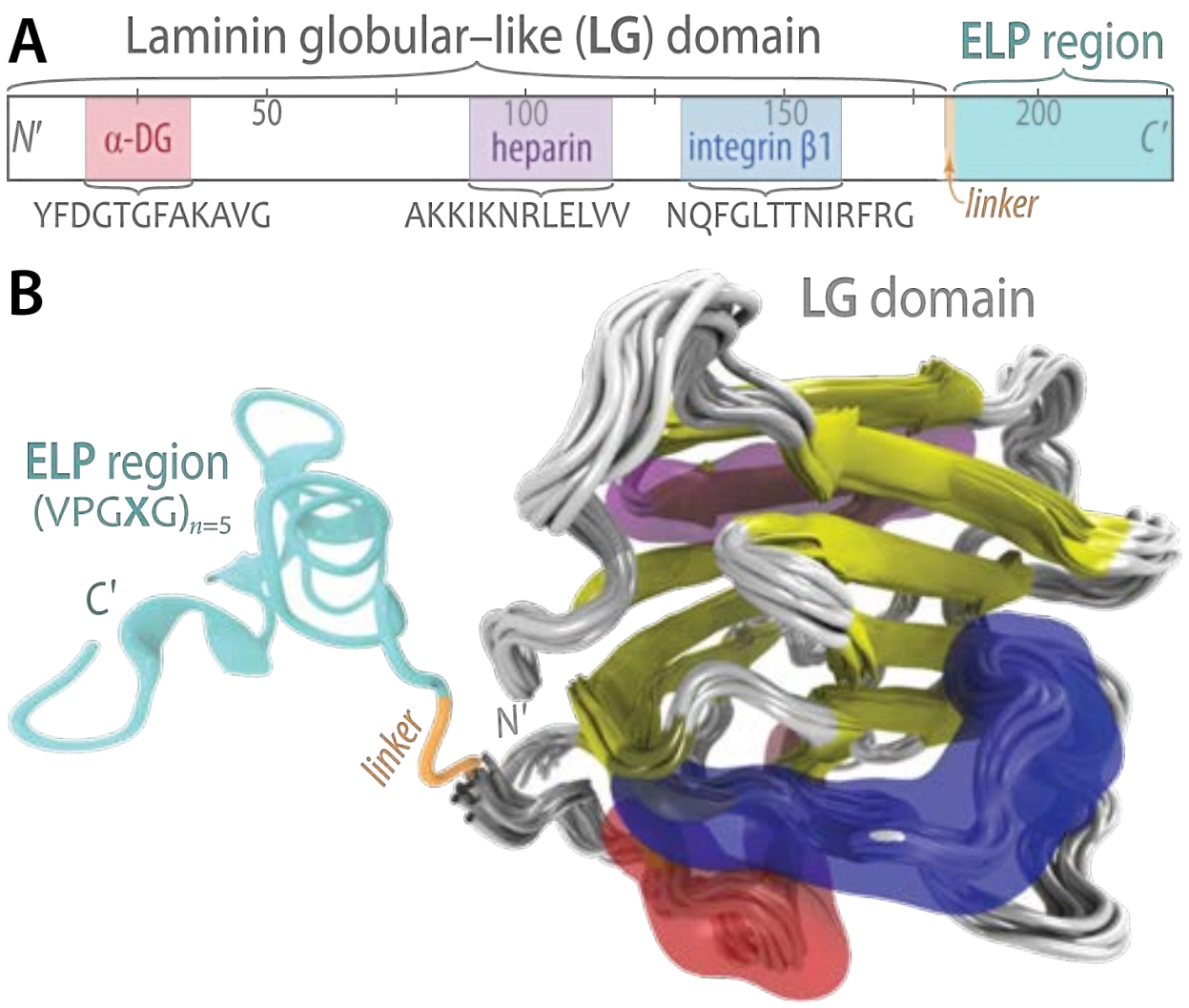

**Figure 1.** The proposed LG–ELP fusion protein. This schematic of our protein engineering design shows a laminin globular-like (LG) domain fused to a C′-terminal elastin-like polypeptide (ELP). (A) The biologically active segments[25, 27] in the LG domain function as recognition/binding sites for α-DG (red), heparin (purple), and integrin-*β*1 (blue). Our ELP repeat region (cyan), consisting of 42 residues of the ELP pentapeptide repeat motif and a three-residue linker (orange), comprises the C′-terminal tail of our fusion construct; this ELP region is intended to act as a self-assembly module. (B) A 3D structural rendition of the fusion protein (ribbon representation) shows the LG domain as an overlay of multiple snapshots from the 100-ns simulation. The LG domain folds as a $\beta$-sandwich, with two sheets (one with six strands and the other with seven strands) stacked atop one another; the colored regions correspond to the recognition sequences in (A). The ELP tail is indicated (cyan), with the specific structure shown here drawn from the 315 K trajectory at $t$ = 1 ns (i.e., after energy minimization and initial trajectory equilibration).

To prepare for simulations under periodic boundary conditions, the initial 3D model of LG-ELP was solvated in a cube of explicit TIP3P water molecules, using the 'solvation box' extension in VMD[62]; a 15-Å padding of solvent, between the solute and nearest box face, was used to mitigate interactions between the protein and its periodic images. Physiological concentrations (150 mM) of $Na^+$ ions, including sufficient $Cl^-$ ions to neutralize the solute's charge, were then added to the solvated system using VMD's 'ionize' plugin. The final simulation cell contained 166,137 atoms, with a cubic box of water measuring 120 Å/edge. The internal energy was minimized for 10,000 steps, and the system was then equilibrated for 10 ns (with a 2-fs integration step) in the NPT ensemble (Figure 2, initial pose). Simulations were conducted over a range of seven temperatures: 290, 295, 300, 305, 310, 315, and 320 K. In each case, temperature and pressure (1 atm) were maintained using a Langevin thermostat and piston. NAMD 2.9 was used for all simulations,[63] with each trajectory extended to a final production time of at least 100 ns. To assess whether trajectory-derived quantities were consistent across our various final (production) runs—and not merely consequences of insufficient/limited sampling—we performed extended simulations. Using the final structure (trajectory frame) from the 310 K simulation (effectively providing a negative control) we computed the corresponding structural quantities of 100-140 ns trajectories at 290, 300, and 320 K. Moreover, we extended the 310 K simulation to 200 ns, as interesting transitions occur near this temperature.

Trajectories were processed and further analyzed using in-house scripts written in the Python[64] and D[65] programming languages, as well as VMD. Root-mean-square deviations (RMSD) for $C^{\alpha}$ atoms were computed with VMD's RMSD extension toolbox. Secondary structures in the ELP region were assigned using STRIDE.[66, 67] Table S1 summarizes all of our LG-ELP–related simulations. All simulation configuration files and analysis scripts are available upon request.

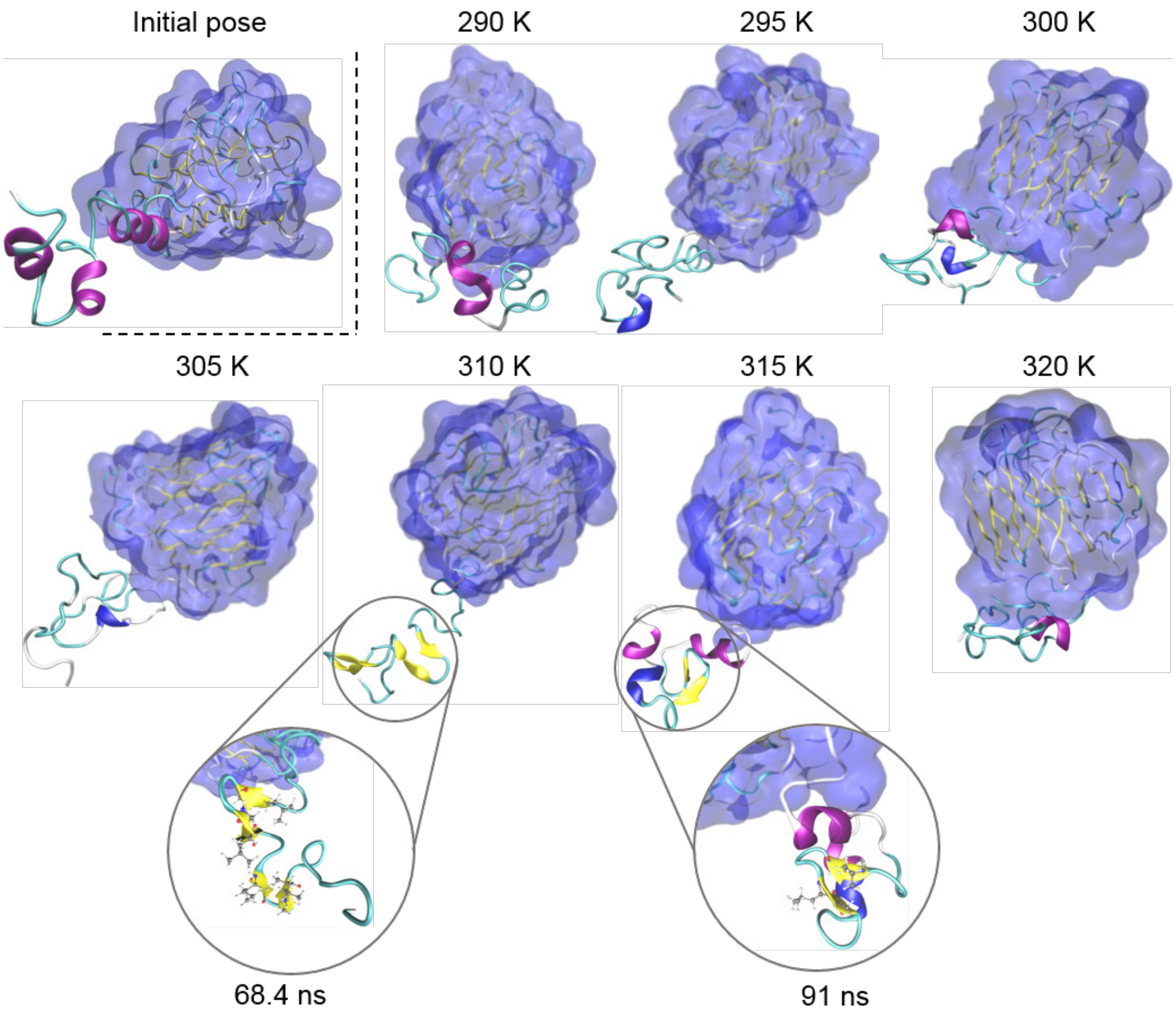

**Figure 2.** Representative structures, illustrating temperature-dependent conformational states of the LG-ELP fusion protein. In the initial pose, the LG-ELP protein is shown immediately after minimization and equilibration of the simulation system, with the LG domain (ribbon diagrams) enclosed by a semi-transparent blue surface. This initial pose was the starting model for simulations at each temperature. The ELP region (ribbons) in this starting state can be seen to be a mixture of helices and coils; the C′-terminus is labelled in this view with α-helices colored purple, $3_{10}$ helices blue, $\beta$-strands yellow, the $\beta$-turn motif cyan and irregular coil regions white. LG-ELP structures are shown from each of the 290–320 K trajectories, with each temperature

indicated and each structural snapshot taken at 100.0 ns. Insets are representative snapshots at 310 K and 315 K, taken from the 68.4 ns and 91 ns timepoints, respectively; the side-chains that contact one another to mediate $\beta$-sheet formation are depicted as ball-and-stick representations (gray carbons, blue nitrogens, red oxygens, and silver hydrogens). These trajectory frames illustrate the formation of $\beta$-sheet regions within the ELP tail.

**Analysis of relative solvent-accessible surface area.** Solvent-accessible surface areas (SASA) were calculated with the SASA tool in VMD, using a standard water probe radius of 1.4 Å. Rost & Sander's method[68] was used to determine the relative solvent accessibility, $RelAcc_i$, of each residue $i$ in the ELP region; this relative accessibility is simply the ordinary accessibility of a residue in a 3D structure ($Acc_i$) normalized by the maximal value possible for that residue type ($RelAcc_i = \frac{Acc_i}{maxAcc_i}$). In our analyses, $RelAcc_i$ values were computed over the entirety of the production trajectories for each simulation temperature.

**Hydrogen-bonding analysis.** Hydrogen bonds were computed using VMD's geometric criteria: namely, a distance cutoff of 3.5 Å and a D-H-A angle cutoff of 30°. Hydrogen bonds between two water molecules were excluded from our calculations. The number of water molecules surrounding the ELP backbone was determined by counting the number of waters within 3.15 Å of the peptide, as previously described.[52] This distance corresponds to the first minimum in the radial distribution function between the oxygen atoms of water molecules and atoms in the peptide backbone (Figure S2).

**Statistical data analysis.** Output data from our Tcl/Tk scripts (used with VMD's Tcl API) were analyzed using tools from the NumPy and SciPy Python packages. Note that all simulations, and subsequent trajectory and data analyses, were of the *full-length* (225–amino acid) LG-ELP protein. In many cases, we show only the ELP region in certain sections of our analyses; this is purely for the sake of clarity and simplicity. Spearman rank-order correlation coefficients, and associated *p*-values, for trajectory-derived data (taken from the beginning to the end of the trajectory)—such as intramolecular hydrogen bonding statistics, the number of neighboring water molecules, etc.—were calculated using SciPy's statistical modules.

## RESULTS AND DISCUSSION

**Temperature-dependent structural transitions of the LG-ELP fusion construct.** To explore the structural properties and conformational dynamics of our model LG-ELP fusion protein (Figure 1) at various temperatures, and illuminate its phase transition behavior, we performed all-atom MD simulations of the protein immersed in a bath of explicit solvent. This system was simulated at temperatures ranging from 290 K to 320 K, with each trajectory extended to at least 100-ns duration. Representative structures from the trajectories show that the ELP region in the initial pose is a mixture of helices and coils, and this region forms more structured $\beta$-strands near 310–315 K (Figure 2). This finding agrees with other studies of the assembly propensity of similar ELP segments (albeit in isolation, not as a fusion partner).[69, 70] We find that the ELP does not associate with the LG domain, and thus the LG domain remains accessible in solution, for binding of bioactive agents such as integrins, heparin, and α-DG. The structural stability and general rigidity of the N′-terminal LG domain is largely maintained throughout each simulation, with the RMSD never exceeding 5 Å (data not shown), as opposed to the far more flexible ELP region (Figure 3).

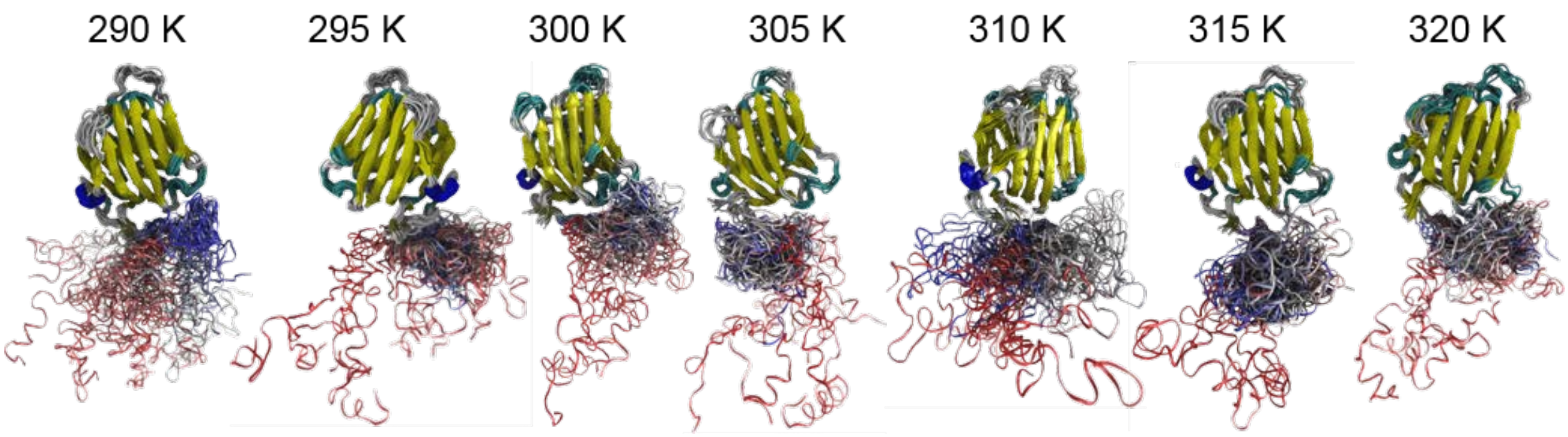

**Figure 3.** Representative structures of the LG-ELP fusion protein simulated at different temperatures. Spatiotemporal evolution of the LG-ELP fusion protein is demonstrated by superimposing frames, taken at 10-ns intervals, from the simulation of the entire fusion protein. The ELP region is colored so as to convey the simulation time, graded from early (red) to middle (gray) to late (blue) timesteps along the MD trajectory. Note the structural rigidity of the LG domain and the conformational flexibility of the ELP region.

As shown by the overlaid structural snapshots in Figure 3, the LG domain's initial structure is largely preserved throughout each simulation. The 'frayed' appearance of the ELP region highlights the structural disorder/flexibility inherent to native elastin-based sequences. At temperatures below 305 K, we see a collapse of the ELP from its initial conformation. A hydrophobic cluster within ELP, towards the end of the 100 ns trajectory, is present for all temperatures except 310 K, where the ELP region becomes extended; this point can also be seen in each contact map (Figure 4). Contact maps are matrices that show, for each residue in a 3D structure, the pairwise distance to all other residues. These symmetric matrices compactly represent the pattern of intramolecular contacts, and in our case reveal a lack of interatomic contacts between the LG and ELP regions (Figure 4). At 310 K, a transient—but noticeable—extension of the ELP occurred, starting at ~75 ns and highlighted by the loss of intra-strand hydrogen bonds (data not shown). This thermally-induced rearrangement of the ELP region may well correspond to the sampling of conformations that would favor higher-order (intermolecular) assembly, and we do not see this structural extension at 315 K (though, as for any simulation, absence of an observation could reflect limited sampling).

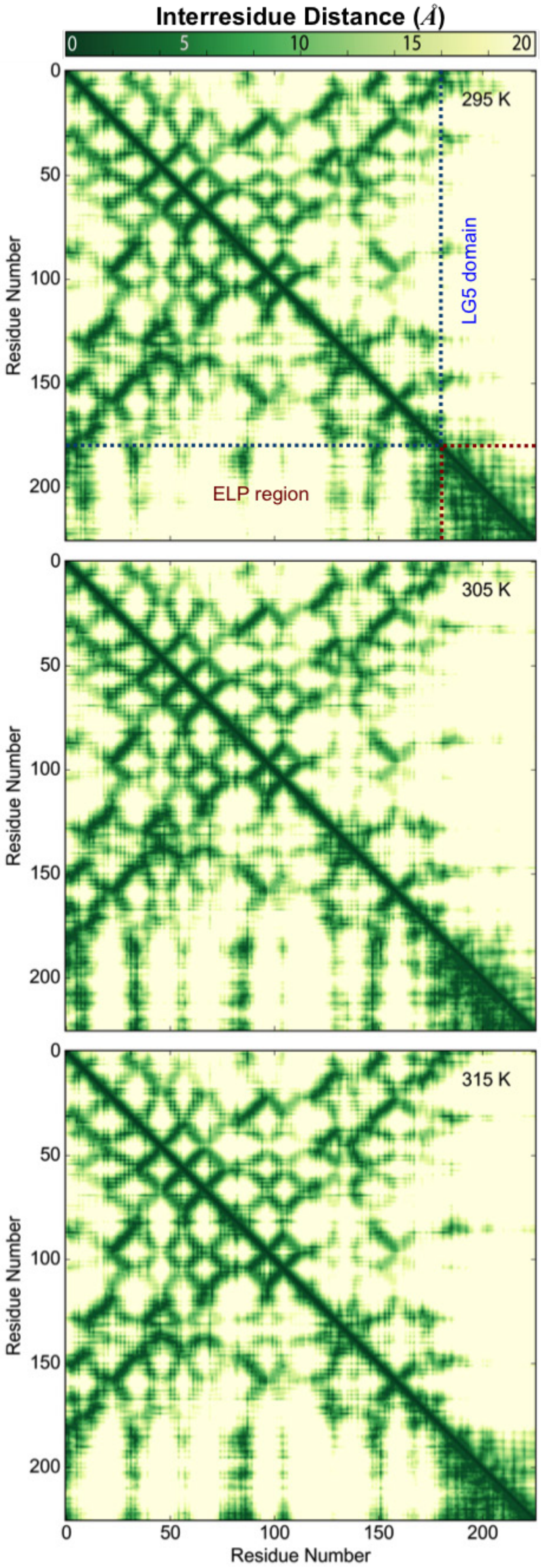
Interresidue Distance (Å)
0
5
10
15
20
295 K
LG5 domain
ELP region
305 K
315 K
Residue Number
0
50
100
150
200

**Figure 4.** Contact maps of the dynamical interactions in our LG-ELP design reveal a lack of persistent LG···ELP interactions, independent of temperature. Contact maps are shown for the full length LG-ELP fusion at the indicated simulation temperatures, with colors graded by the pairwise distance (scale bar) between the two side-chains under consideration. The two discrete structural units in our fusion, i.e. the LG domain and ELP region, are demarcated by blue and red lines, respectively (for clarity, this is drawn only in the 295 K map). The classic crisscross patterns, highlighted by stripes of contacts perpendicular to the main diagonal, are indicative of the *β*-sheet core of the LG domain. Because an ordinary (symmetric) contact map contains two-fold redundant information, here we show (i) the *minimum* inter-residue distance in the lower triangular matrix, and (ii) the *mean* inter-residue distance, averaged over an entire trajectory, in the upper triangle. At all simulation temperatures, no stably persistent intra-molecular contacts (short distances) are found between the LG and ELP regions, as illustrated by (i) the high-intensity (short-distance) square submatrices at the lower-right of each map, indicating that most ELP residues interact with other ELP residues (not LG residues), and (ii) the vertical white stripes toward the right of each matrix, indicating a dearth of LG···ELP contacts. Thus, the ELP polypeptide does not engage in spurious/unwanted interactions with the LG region in solution, at least not on the 100-ns timescale of these simulations. (Contact maps for all simulated temperatures can be found in Figure S8.)

**Secondary structure composition and its dependence on temperature.** We examined the structural transitions from the initial starting peptide structure to the final conformational ensemble, focusing on the ELP region of the LG-ELP fusion. At all temperatures, the ELP region exhibits a significant amount of unstructured character (*β*-turn and 'other' in STRIDE), with these two classes accounting for most of the secondary structures in the ELP (Figures 5, S3, and S4). These findings are consistent with solid-state NMR data[59, 71] and CD spectroscopy[72, 73] of similar ELP sequences, where residues within the pentapeptide repeat preferentially adopt *β*-turn structures. We found that the ELP region accrues *β*-strand character over the course of a 100-ns trajectory at physiologically-relevant temperatures (Figure 5, Video S1), and we posit that this *β*-strand enrichment can serve as a useful structural property for achieving temperature-triggered LG-ELP assembly; such assembly can occur via intermolecular *β*···*β*-strand contacts, e.g. by the domain swapping mode of *β*-rich protein association.[74, 75]

In simulations at 305 K, there is a sharp reduction of α-helicity, followed by a complete loss of helical structure after 74 ns (Figure S3 and S5). The secondary structure distribution at 305 K also shows a bimodal distribution in *β*-turn and 'other' motifs (Figure 5), indicating the preferential sampling of these two discrete conformational states. At 310 and 315 K, there is an increase in *β*-sheet character. The occurrence of *β*-sheet–like structures at temperatures above the phase transition has been experimentally detected in similar, single-molecule ELP systems.[46, 52, 73, 76, 77] The drastic change in secondary structural content found in our trajectories suggests that heating the system potentially destabilizes polyproline-induced α-helix conformations, perhaps by selectively decreasing the stability of water solvation effects[78, 79]. Such a disruption in helical propensity is consistent with the findings of Li et al.[46] and Ohgo et al.,[59] where, at higher temperatures, the proline in (VPGXG) adopts torsion angles similar to type-I and type-II *β*-turns. This shift in secondary structure in our LG-ELP system is especially prominent at 320 K, where there is a complete loss of *β*-sheet character, and the reduction of *β*-bridges with respect to 310 K and 315 K is associated with the increase in *β*-turns within the system. At lower temperatures, the composition favors more α-helical and 'other' secondary structures ($3_{10}$ helices, π-helices, random coils, etc.). The pattern of sampling that we find in secondary structure formation, as a function of temperature, suggests that 310 K is near the target temperature at which macromolecular ordering of the LG-ELP fusion may occur.

This phenomenon associated with the structural changes accompanied by the phase transition is further demonstrated by the distribution of secondary structural content in the 100-200-ns trajectory. The time evolution of the secondary structure profile in the extended simulation at 310 K showed four distinct regions of persistent □-sheet like conformations, Leu4-Gly5$_{200-201}$ - Leu4-Gly5$_{205-206}$ and Ile4-Gly5$_{210-211}$ - Ile4-Gly5$_{214-215}$ (Fig. S6, Video S4) with reduced conformational flexibility. Using the final trajectory frame of 310 K as a starting structure, we extended the simulation from 100 ns to 140 ns at 290 K, to assess the potential artefacts of limited sampling of structural classes. Reassuringly, we found that the □-sheet state does not persist, and in fact it disappears within 5 ns (Fig. S4, S7, Video S2). Similarly, a transition from the 310 K trajectory to 320 K corresponds to a decrease of □-sheet content. At 300 K, however, the temperature shift resulted in a seemingly stable, extended □ conformation of the peptide backbone in the Gly5$_{201}$-Leu4$_{205}$ region from 100-140 ns (Fig. S4 and S6, Video S3). This result indicates that the intramolecular contacts between these nonpolar side-chains might be attributable to a population of pre-existing conformations from the previous structural ensemble at 310 K, as these are precisely the same □-sheet forming residues from the initial 100-ns trajectory.

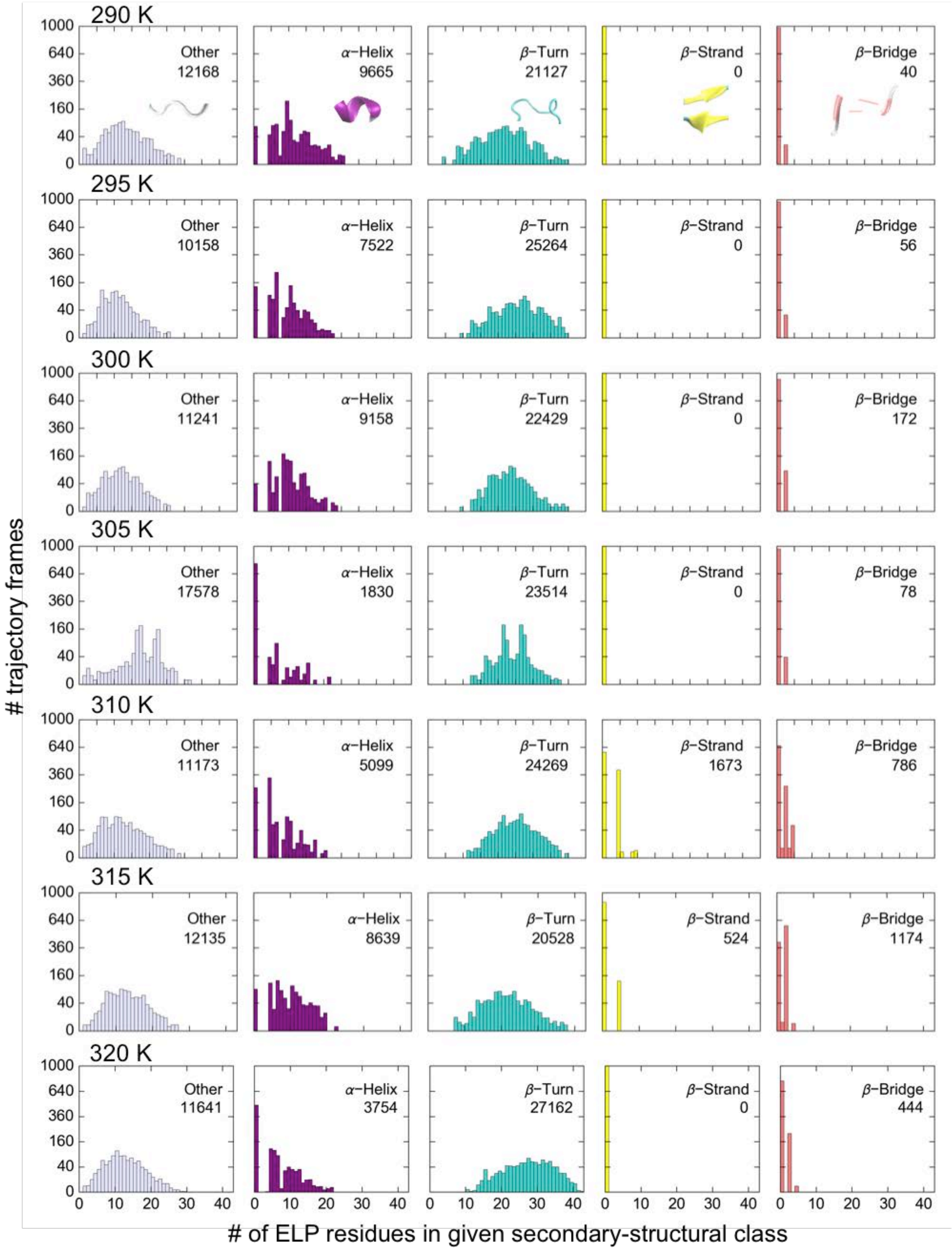
Other
12168
10158
11241
17578
11173
12135
11641
α−Helix
9665
7522
9158
1830
5099
8639
3754
β−Turn
21127
25264
22429
23514
24269
20528
27162
β−Strand
0
0
0
0
1673
524
0
β−Bridge
40
56
172
78
786
1174
444
# trajectory frames
# of ELP residues in given secondary-structural class

**Figure 5.** Secondary structural content of the ELP region as a function of temperature across the 290–320 K series. For simplicity, these trajectory analysis results are shown only for the ELP region, instead of the full-length LG-ELP fusion; there are no noticeable changes in the structural content of the LG domain in all of our simulations. These secondary structure analyses show that the average conformational behavior of a single ELP monomer strongly depends on simulation temperature. Numbers written as insets within each panel give the total number of times that the secondary structure was detected in the simulation. Cartoon representations, shown as secondary structure thumbnail schematics in the first row, match the colors in the histogram. The predominant conformations exhibited by the ELP are $\beta$-turns and 'other' structures. At low temperatures, α-helical and $\beta$-turn structures are prevalent, with minimal $\beta$-strand and bridge structures. However, states with greater $\beta$-sheet structural content occur as the temperature goes from 305 K to 310 K, indicating a possible order/disorder phase transition. Additionally, a significant shift in the character of the $\beta$ structure, from strand to bridge, occurs at 315 K. The complete lack of $\beta$-strand structure at 320 K and subsequent rise in $\beta$-turns corresponds to an increased flexibility of the ELP backbone at higher temperatures.

**Relative solvent accessible surface area and visualization of association interactions.** Conformational transitions can be analyzed via dynamical correlation functions, which provide information on how a molecule can interact with the surrounding solvent. We evaluated the solvent accessible surface area (SASA) of the ELP region in order to characterize the local ordering and solvation dynamics of the system. The SASA can help quantify protein surface–water contacts, and it is a parameter that has long been associated with the thermodynamics of protein structure, as related to the hydrophobic effect and folding.[80]

We find no strong trend in solvent exposure properties for residues in the ELP region (Figure 6). For all residues, a linear regression of SASA against temperature yields fits with $R^2$ values less than 0.5 (data not shown). This result suggests that the structural transitions of ELP regions do not strongly correlate with the SASA of any specific residue, representing a notable departure from previous models of ELP phase transitions.[69, 81] There is also a striking lack of correlation between $RelAcc_i$ and temperature. Linear regression gives $R^2$ values less than 0.35 for each residue, again suggesting that any ELP phase transition in this temperature regime is not accompanied by gross structural rearrangements. While the hydrophobic regions of ELP have been thought to become more exposed at elevated temperatures (at least for isolated ELP segments, unfused to other proteins),[82] our simulations do not reveal any such correlation. Though the *RelAcc* of our ELP residues is uncorrelated with temperature, the values do fluctuate (Figure 6), and no single residue is consistently buried or consistently exposed. The ELP phase transition, therefore, seems to be marked most strongly by the formation of $\beta$-sheet secondary structures, without any concomitant gross structural rearrangements (at least in terms of SASA).

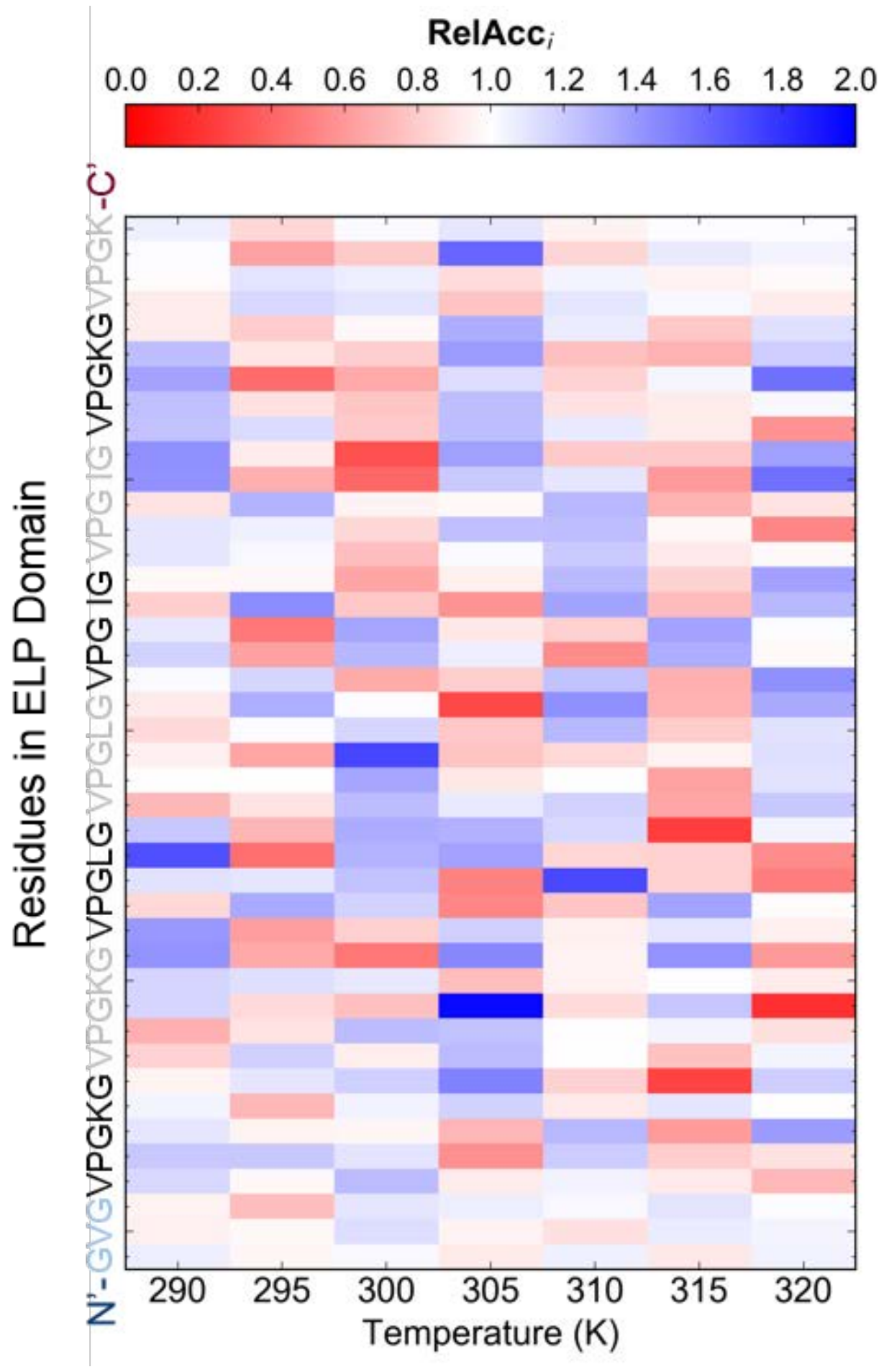

**Figure 6.** Temperature-dependent changes in relative SASAs of individual residues in the ELP region. The relative solvent accessibility, *RelAcc*$_i$, represents the accessible surface area of a residue in the context of a (potentially folded) polypeptide. Blue colors indicate that a residue is more solvent-exposed than average, while red indicates that a residue is more buried than it otherwise would be (outside the context of the peptide).

A close examination of the intramolecular contacts, i.e. within the fusion protein, reveals that the formation of $\beta$-sheets by ELP residues is not occluded or otherwise hindered by interatomic contacts between the ELP region and the LG domain (Figures 4, 5, and S8). From a protein design perspective, this is most reassuring: our simulations suggest that the ELP region will be accessible in solution, free of significant interactions with the nearby LG domain. Similarly, the LG domain's function should not be abrogated by the presence of ELP, and we expect putative ELP···ELP interactions to mirror those found in previous studies of ELP aggregation.[69, 73]

**The role of hydration in compact conformations.** We investigated the time-dependent hydration properties of our fusion's ELP region by examining the intramolecular hydrogen bonding (within ELP) and the number of water molecules hydrogen-bonded with the ELP. The number of surrounding water molecules decreases and the number of intra-peptide hydrogen bonds increases, with increasing temperature from 305 to 315 K, and then a dip occurs at 320 K (Figure 7). There is a slow decrease, with time, in the number of solvating water molecules at all simulated temperatures (Figure S9A). The 310 K trajectory features an intriguingly abrupt dip in the number of water molecules at 64 ns and at 82 ns. The displacement of water molecules with higher temperatures is consistent with a model wherein desolvation (e.g., of nonpolar side-chains) biases specific (e.g., polar) segments of the amphipathic ELP chain into more compact conformations, such as $\beta$-turns and strand-like conformations. Helical structures are often unfavorable at elevated temperatures for entropic reasons, such as a greater loss, upon folding, of orientational and other conformational degrees of freedom[83, 84]. Thus, higher temperatures may indirectly, via effects on solvation structure, enhance the stability of $\beta$-sheet formation in relatively disordered conformational ensembles, such as that of ELP. Changes in hydration density exhibit a correlation with $\beta$-sheet propensity along all trajectories (Figures S5 and S9B). A Spearman's rank-order correlation coefficient of –0.83 ($p$ = 0.04, for 100 ns) indicates a moderately strong negative correlation between intramolecular hydrogen bonding and surrounding water molecules with increasing temperature. This quantity captures the fact that, at elevated temperatures, the ELP region preferentially contacts itself rather than water—indicative of a phase transition[59]. The increased number of hydrogen bonds above 305 K suggests a coil-to-globule transition.[76] A possible model is that, at high temperatures, insufficient conformational order exists to allow for formation of a single, well-defined structural state. As such, at lower temperatures the increased rigidity of the system would not facilitate the formation of intramolecular peptide···peptide hydrogen bonds, which would, instead, be replaced by inter-molecular hydrogen bonds with the surrounding water structure.

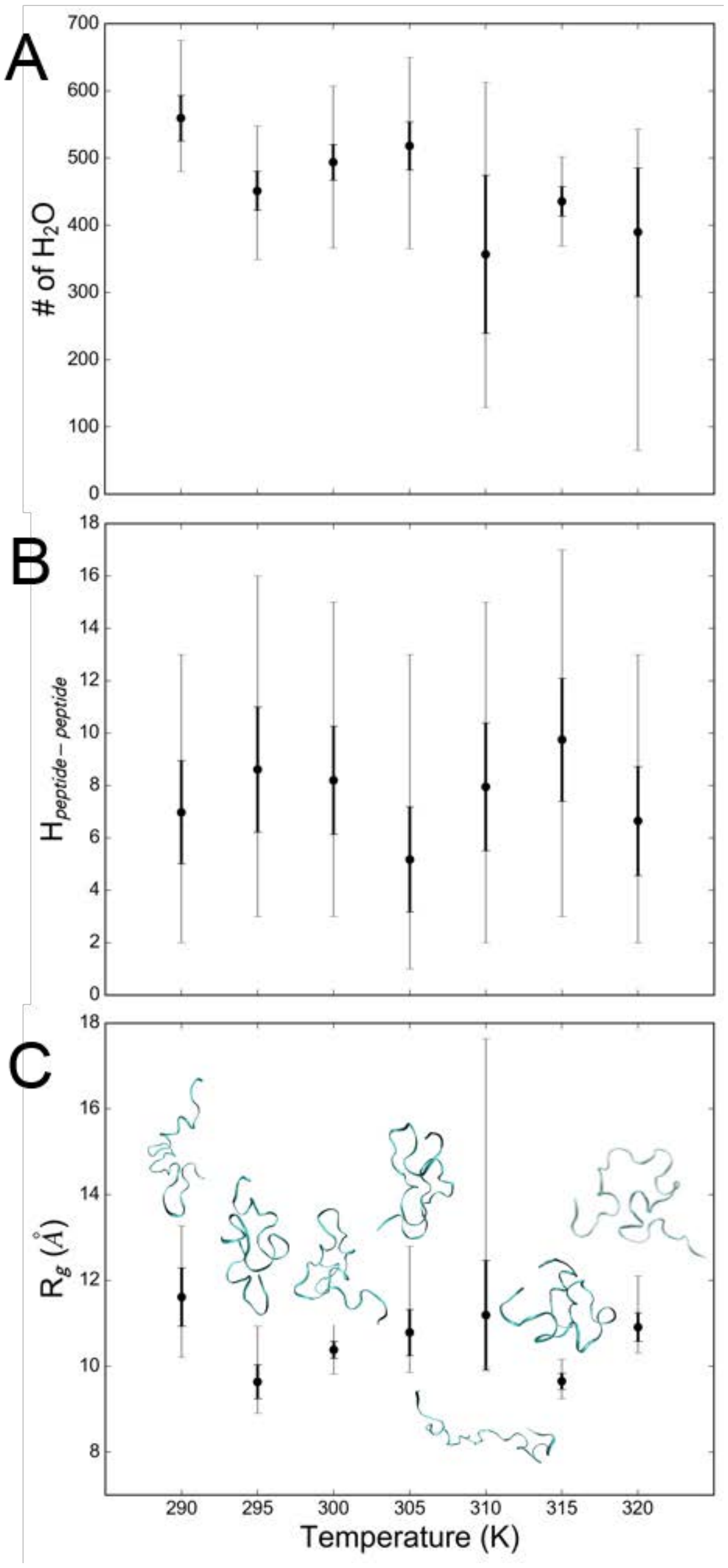

**Figure 7.** Changes in degree of hydration, hydrogen-bonding, and overall structure of the ELP region. (A) The number of water molecules is counted within 3.15 Å of the ELP, with varying temperature. An abrupt decrease in the number of surrounding water molecules suggests that this change is associated with the formation of $\beta$-sheets at 310 K in the ELP region. (B) The number of intramolecular peptide···peptide hydrogen bonds ($H_{peptide\text{-}peptide}$). The formation of intramolecular hydrogen bonding has been observed for many peptide aggregates that exhibit

LCST behavior; however, the large number of disordered conformational states of our ELP hinders us from discerning any trend as regards a temperature that might be indicative of a phase transition. (C) Temperature dependence of the radius of gyration ($R_g$). Proteins in all simulated temperatures exhibit temperature-induced collapse (relative to the initial starting structure). Only 310 K and 320 K show a slight expansion of the polypeptide chain, while all other proteins exhibit compaction—reminiscent of the 'hydrophobic collapse' in typical (water-soluble) globular proteins. In all panels, black error bars represent standard deviations and gray error bars show min/max values. Only the last 40 ns of the trajectories at each temperature are included in the analysis shown here.

Coupled protein···solvent interactions are a key element of a system's structural properties and dynamical behavior in any order/disorder transition (e.g., protein folding), but time-resolved experimental data on such interactions are not easily obtained, at least at high spatial resolution. Atomistic simulations can provide information about literally each inter-atomic contact, including the dynamical networks of (i) apolar interactions within a protein, (ii) protein···solvent contacts, and (iii) solvent···solvent contacts, all of which are important factors in macromolecular folding and binding. The compactness of a biomolecular 3D structure—and, by inference, the degree of formation of a hydrophobic 'core'—can be measured as the radius of gyration, $R_g$. The time-evolution of $R_g$ for the ELP region alone (Figures 7C, S10, and S11) does not clearly reveal a sharp phase transition, unlike many biopolymers that exhibit LCST behavior.[43, 77, 85] Though $R_g$ data are, in principle, experimentally accessible via solution-state measurements, e.g. Guinier analysis of small-angle X-ray scattering data,[86] such approaches to extracting $R_g$ values are confounded by phase changes in going from a soluble to insoluble state, as is common with many polymers that demonstrate LCST behavior. Our simulations reveal that the ELP portion of our fusion protein adopts $\beta$-strand secondary structures at high temperatures, implying that this region can undergo structural changes, akin to order/disorder phase transitions, and form ordered complexes. Intriguingly, the drastic solute re-structuring that is often associated with LCST behavior[82] does not appear to be a feature in our system's transition. At higher temperatures, the unfolding or 'elongation' of the polypeptide (Figure S10 and S11) is primarily entropically driven, but at a critical temperature (near 315 K in our system), the chain collapses because the loss of configurational entropy of the side-chains and backbone is counterbalanced by entropic changes in the network of solvent···{solvent, protein} interactions.[87, 88] To assess whether our findings were consistent with our results from the first 100-ns trajectories, we performed additional simulations at 290 K, 300 K, 310 K, and 320 K using the final (100 ns) frame from the 310 K simulation as the starting structure for each different temperature. These extended trajectory data support the argument of a structural transition near 310 K, where it is represented by a gross structural rearrangement of the polypeptide backbone. This transient state is characterized by a 'two-state' equilibrium between the collapsed and extended conformation (Fig. S11) within the ELP region. At low temperatures, i.e. 290 K and 300 K, we continue to observe a collapsed state, which is stabilized by the relatively strong peptide···peptide and peptide···water interactions, compared to the extended conformation at 310 K.

As a final step of analysis, we considered the 'end–to–end' distance, taken as the simple Euclidean distance between the N′- and C′-termini of a given polypeptide segment, as another geometric measure of peptide compactness. Monitoring the dynamics of the end–to–end distance for the ELP region (Figure S12) revealed that this part of our fusion design can explore a substantial region of conformational space without altering its global shape (as indicated by a relatively constant $R_g$ value). Note that this behavior differs from that of larger, 'ordinary' globular proteins, where the detailed 3D structural changes that correspond to transitions between nearby local minima on the energy landscape effectively act as barriers to the rapid sampling of conformational space, thereby decreasing kinetic rates of transitions.[89-91]

## Conclusions

Classical, all-atom MD simulations were used to examine the structural properties and conformational dynamics of an engineered, laminin–mimetic elastin–like fusion protein, referred to here as LG-ELP. Analyses of the temperature-dependent conformational changes in full-length LG-ELP—in terms of secondary structural content, solvent accessible surface area, hydrogen bonding, and hydration properties—illuminate the phase transition behavior of this fusion protein. The increased structuring of the protein, and the opportunity that that presents for engineering noncovalent interactions, provide a platform for the rational design of macroscopic material properties[92]. The secondary structural elements in a peptide are known to correlate with the compliance, stiffness, density, and other mechanical properties of hydrogels built upon the given peptide.[93, 94] In this work, we computed atomically-detailed MD trajectories of an engineered LG-ELP protein design at several temperatures thereby providing us with an *a priori* view of the phase behavior of our design as a function of temperature in the physiological range; reassuringly, we found that the ELP region of our fusion protein did not engage in spurious interactions with the LG domain. This type of information is invaluable in guiding the design of new fusion protein sequences and motifs with desired biological functionalities. Ultimately, our strategy can be used to simulate multiple fusion protein designs, rank-order them, and synthesize those candidates that exhibit the desired phase transition behavior. Because our strategy of using simulations is physics-based, our approach also illuminates the secondary and tertiary structural properties of our LG-ELP fusion, as well as physicochemical properties such as the coupled dynamics of the solvation environment and its influence on the phase transition behavior of our design.

Simulations are enjoying increased use in the analysis of protein structure and function, but to our knowledge an MD-based simulation methodology has not been used in the manner reported here: namely, to help guide the design and iterative refinement of novel fusion proteins that can act as stimuli-responsive cellular matrix materials. There exist relatively few examples of long-time, all-atom simulations of polypeptide-based materials. The simulations reported here elucidate the relationships between solvation, hydrophobicity, structural dynamics and other atomically-detailed properties, for a novel biomolecular system, and our strategy offers a robust and extensible platform to guide future design and syntheses of protein biomaterials. In particular, our general computational approach can be readily applied in the rational design of engineered extracellular matrix proteins for constructing stimuli-responsive and biocompatible materials for applications in drug delivery, tissue engineering, and regenerative medicine.

## ASSOCIATED CONTENT

**Supporting Information**. The initial starting structure of the LG-ELP protein, the radial distribution function (RDF) of oxygen atoms surrounding the ELP backbone, as well as the time evolution of surrounding water molecules, time evolution of the secondary structural profile for extended simulations, hydrogen bonds, $R_g$, end-to-end distances, simulation videos, and secondary structural content as alluded to in the main text. This material is available free of charge via the Internet at http://pubs.acs.org.

AUTHOR INFORMATION

**Corresponding Author**

*Email: lampe@virginia.edu

**Notes**

The authors declare no competing financial interest.

ACKNOWLEDGEMENTS: We thank K. Holcomb & A. Munro (UVa) of UVa's *Advanced Research Computing Services* for exceptional computer support. Computations were performed on UVa's high-performance cluster, *Rivanna*, with financial support provided by the School of Engineering & Applied Sciences, Data Science Institute, and College of Arts & Sciences. Portions of this work were supported by UVa (KJL), the Jeffress Memorial Trust grant J-971 (CM), Jeffress Memorial and Carman Trust grant 2016.Jeffress.Carman.7644 (KJL) and NSF CAREER award MCB-1350957 (CM).

Table of Contents Graphic and Manuscript Synopsis:

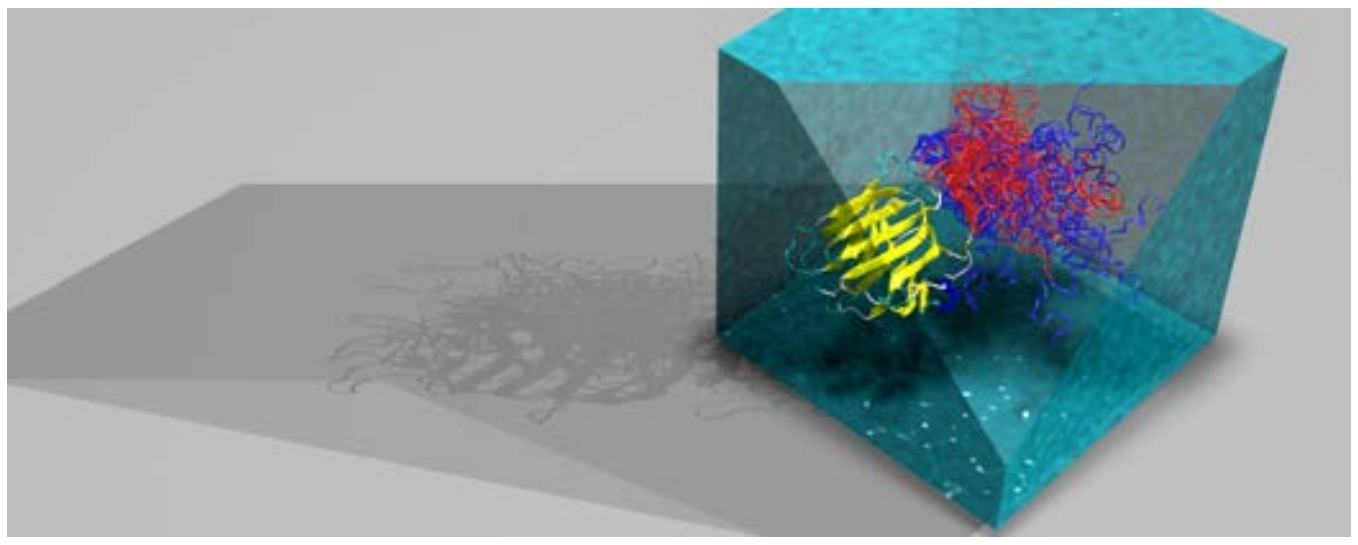

All-atom molecular dynamics simulations of a designer laminin–elastin fusion protein in explicit solvent reveal temperature-dependent conformational changes, in terms of secondary structure composition, solvent accessible surface area, hydrogen bonding, and surface hydration. These properties illuminate the phase behavior of this fusion protein, via the emergence of $\beta$-sheet character in physiologically-relevant temperature ranges.

# Supporting Information

Toward a Designable Extracellular Matrix: Molecular Dynamics Simulations of an Engineered Laminin-mimetic, Elastin-like Fusion Protein

James D. Tang[1], Charles E. McAnany[2], Cameron Mura[2], Kyle J. Lampe[1*]

[1]Department of Chemical Engineering, University of Virginia, Charlottesville, VA 22904, USA

[2]Department of Chemistry, University of Virginia, Charlottesville, VA 22904, USA

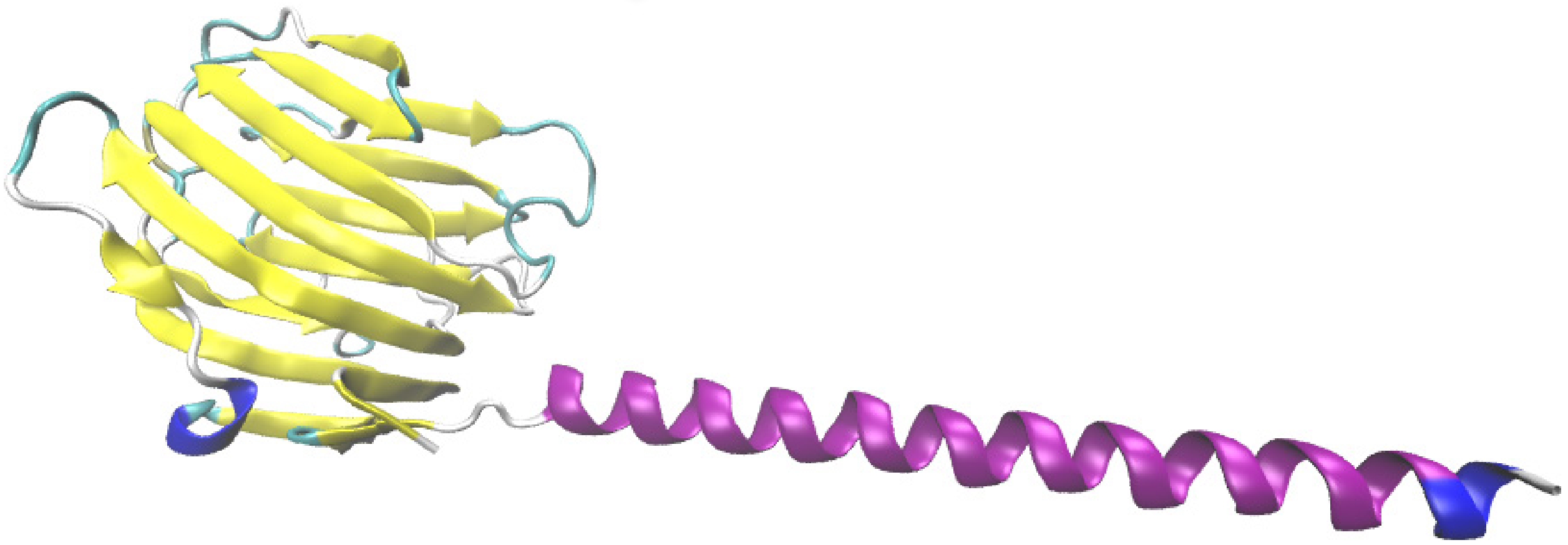

Figure S1: Initial starting structure of the LG-ELP protein. The LG5 domain was drawn from the crystal structure of the mouse homolog of the laminin α2 chain (1DYK), while the ELP domain was built using Avogadro's peptide builder, and was modelled as a canonical α-helix starting structure, with backbone torsion angles $\varphi = -60°$, $\psi = -40$.

**Table S1**. Summary of MD simulation systems of the engineered LG-ELP fusion protein.

| System name | Trajectory Duration (ns) |
|---|---|
| LG5-ELP[K2L2I2K2] | Solvation, Minimization |
| LG5-ELP[K2L2I2K2] | Equilibration 10ns |
| LG5-ELP[K2L2I2K2] 290K | 100 |
| LG5-ELP[K2L2I2K2] 295K | 100 |
| LG5-ELP[K2L2I2K2] 300K | 100 |
| LG5-ELP[K2L2I2K2] 310K | 100 |
| LG5-ELP[K2L2I2K2] 315K | 100 |
| LG5-ELP[K2L2I2K2] 305K | 100 |
| LG5-ELP[K2L2I2K2] 320K | 100 |
| LG5-ELP[K2L2I2K2] 310K final trajectory to 290 K | 40 |
| LG5-ELP[K2L2I2K2] 310K final trajectory to 300 K | 40 |
| LG5-ELP[K2L2I2K2] 310K final trajectory to 320 K | 40 |
| LG5-ELP[K2L2I2K2] 310K trajectory extension to 200 ns | 100 |

Table S1: Atomistic MD simulations were performed using the NAMD 2.9 code and the CHARMM36 force-field for the protein system. The protein was solvated in explicit water with periodic boundary conditions and simulated as described in the *Methods* section.

Tang et al. (2016) Figure S2

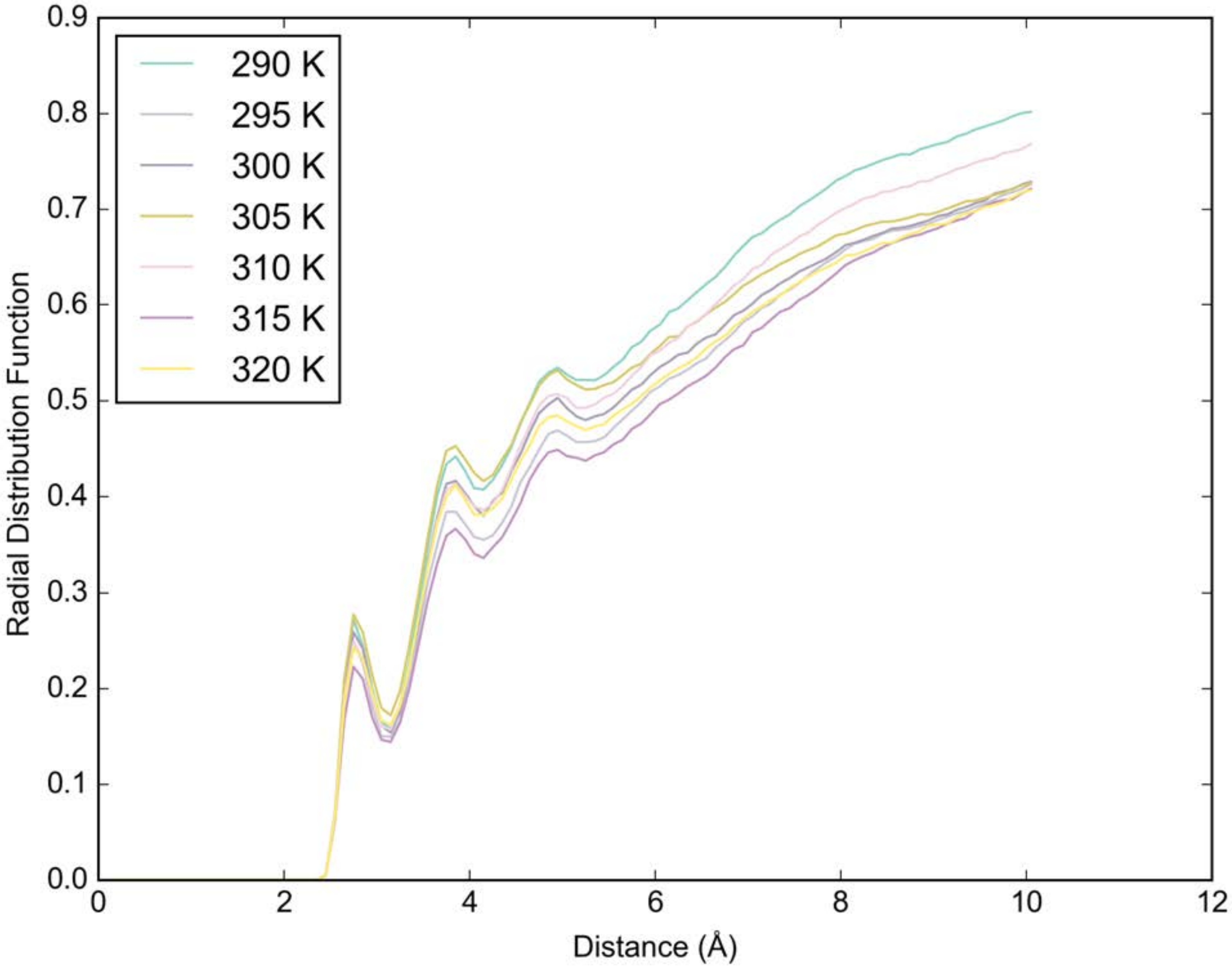

Figure S2: Radial distribution function (RDF) of oxygen atoms around the elastin-like polypeptide (ELP) backbone. The RDF is plotted as a function of temperature, and the first hydration shell was chosen to be the minimum for subsequent analysis in determining the number of surrounding water molecules (Figure 6a).

Tang et al. (2016) Figure S3

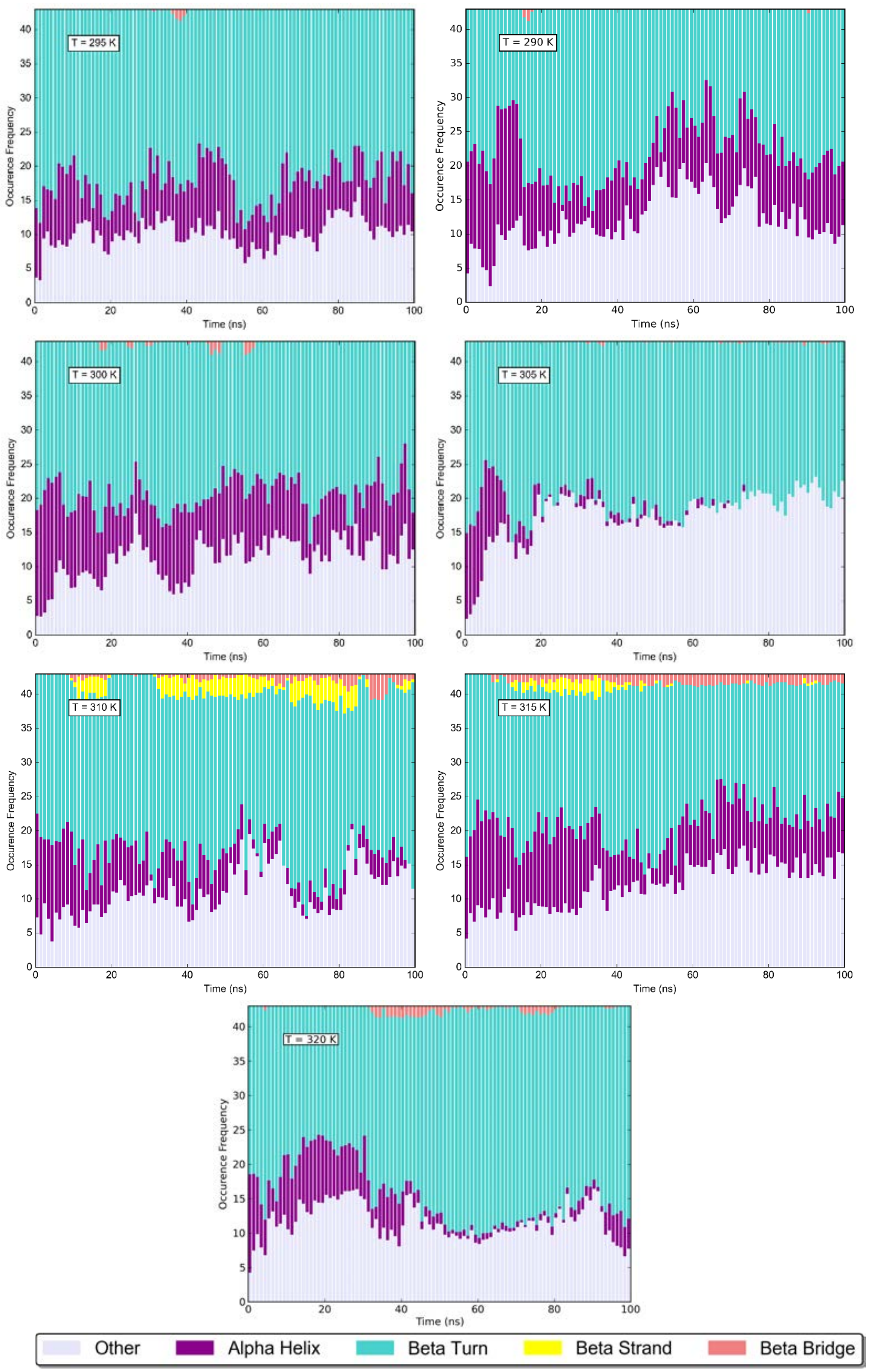

Figure S3: Secondary structural content across a range of temperatures as a function of time. Simulated systems were sampled at different temperatures (five degree increments from 290 K to 315 K).

Tang et al. (2016) Figure S4

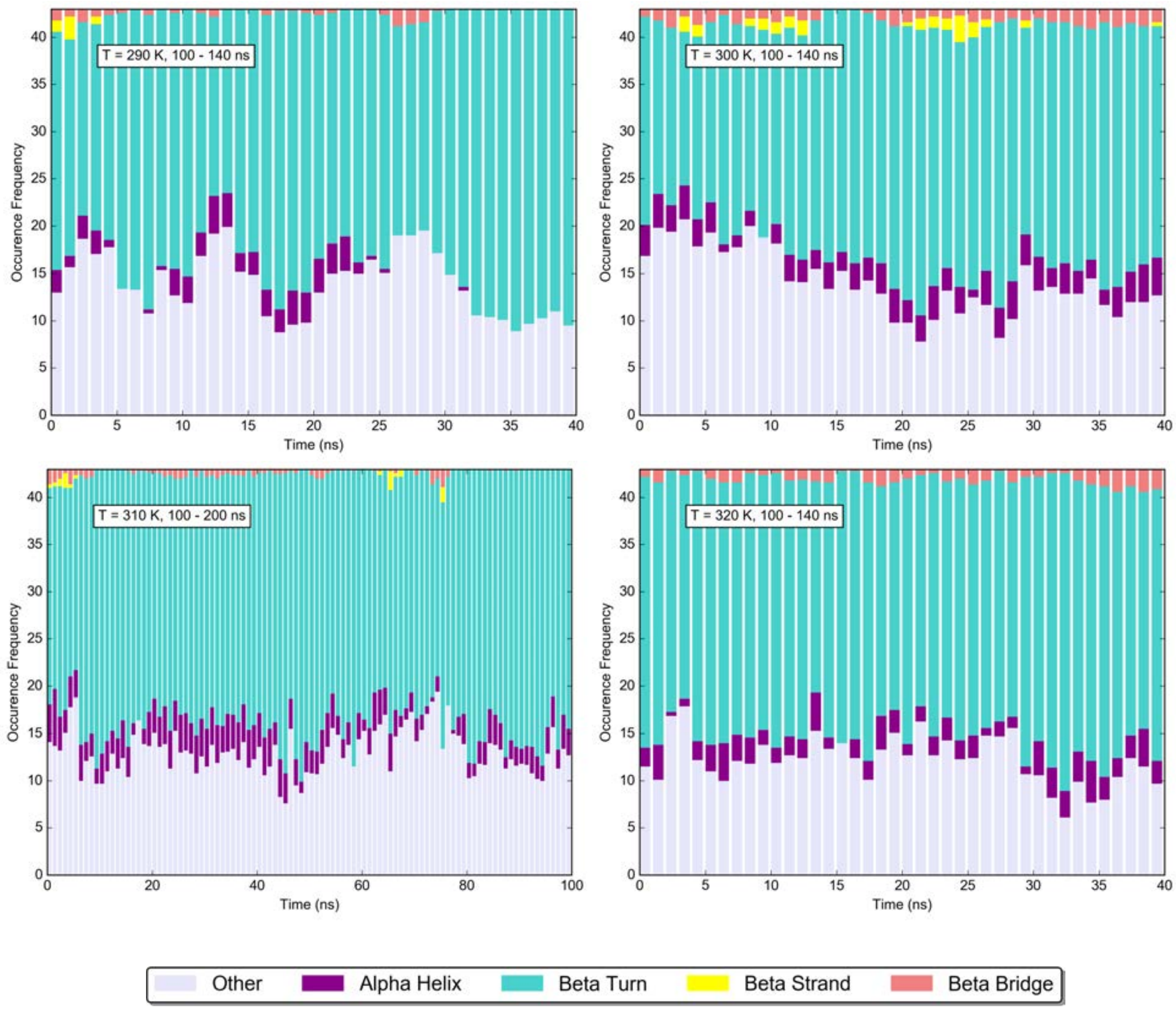

Figure S4: Secondary structural content across a range of temperatures as a function of time for extended simulations (100 – 140 ns for 290 K, 300 K, and 320 K and 100 – 200 ns for 310 K)

Tang et al. (2016) Figure S5

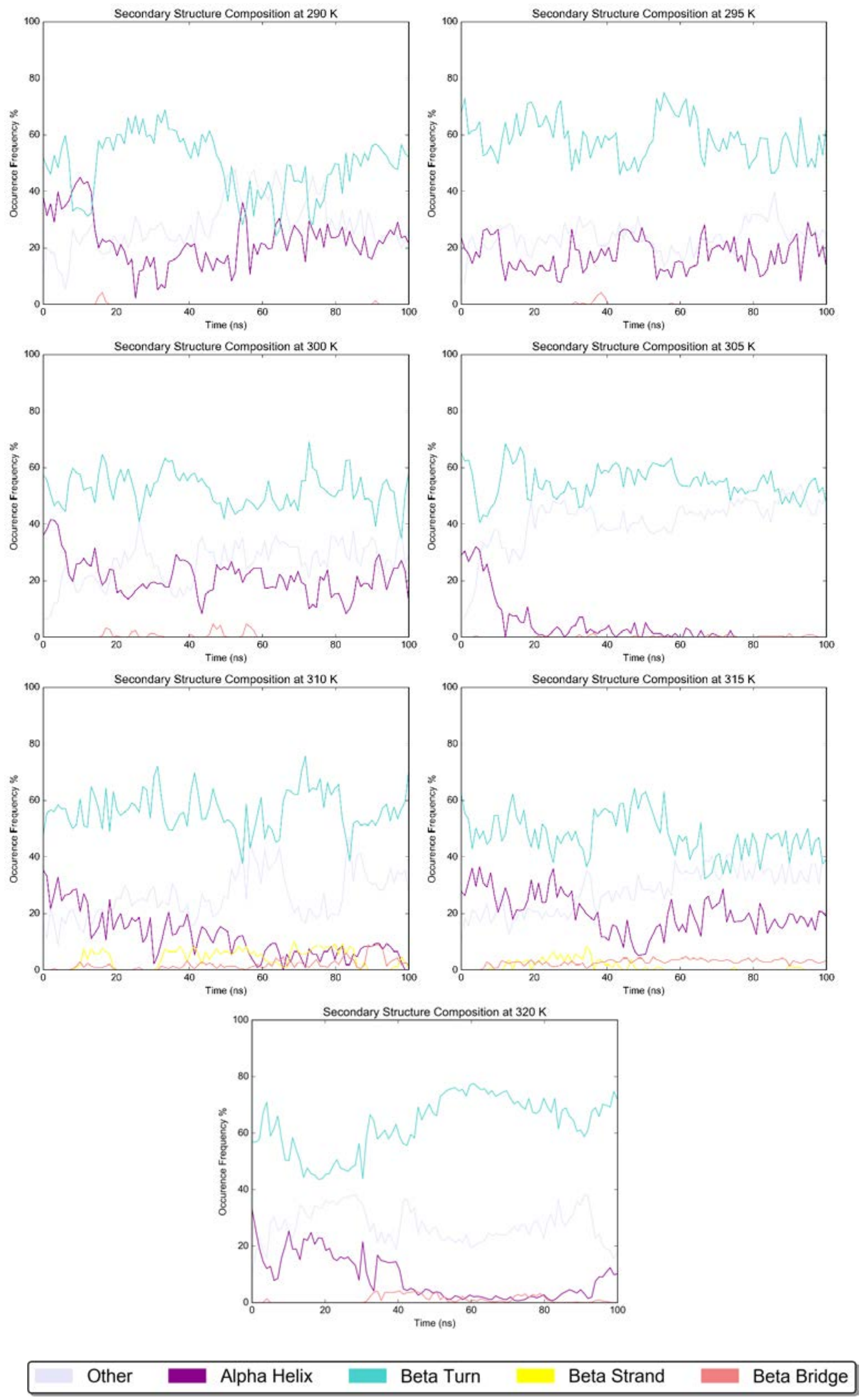

Figure S5: Frequency of occurrence for secondary structural content across a range of temperatures as a function of time.

Tang et al. (2016) Figure S6

(a)

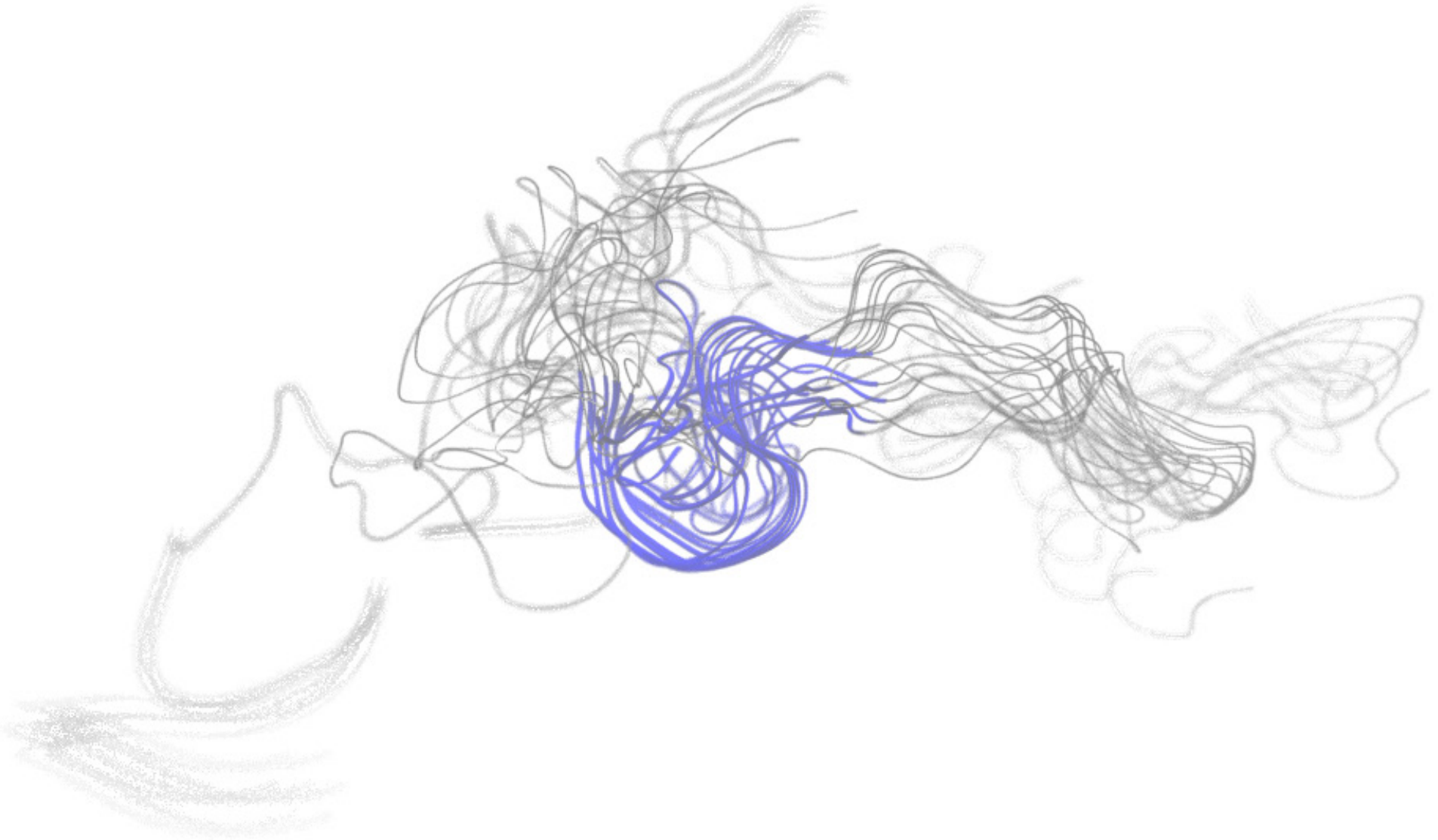

(b)

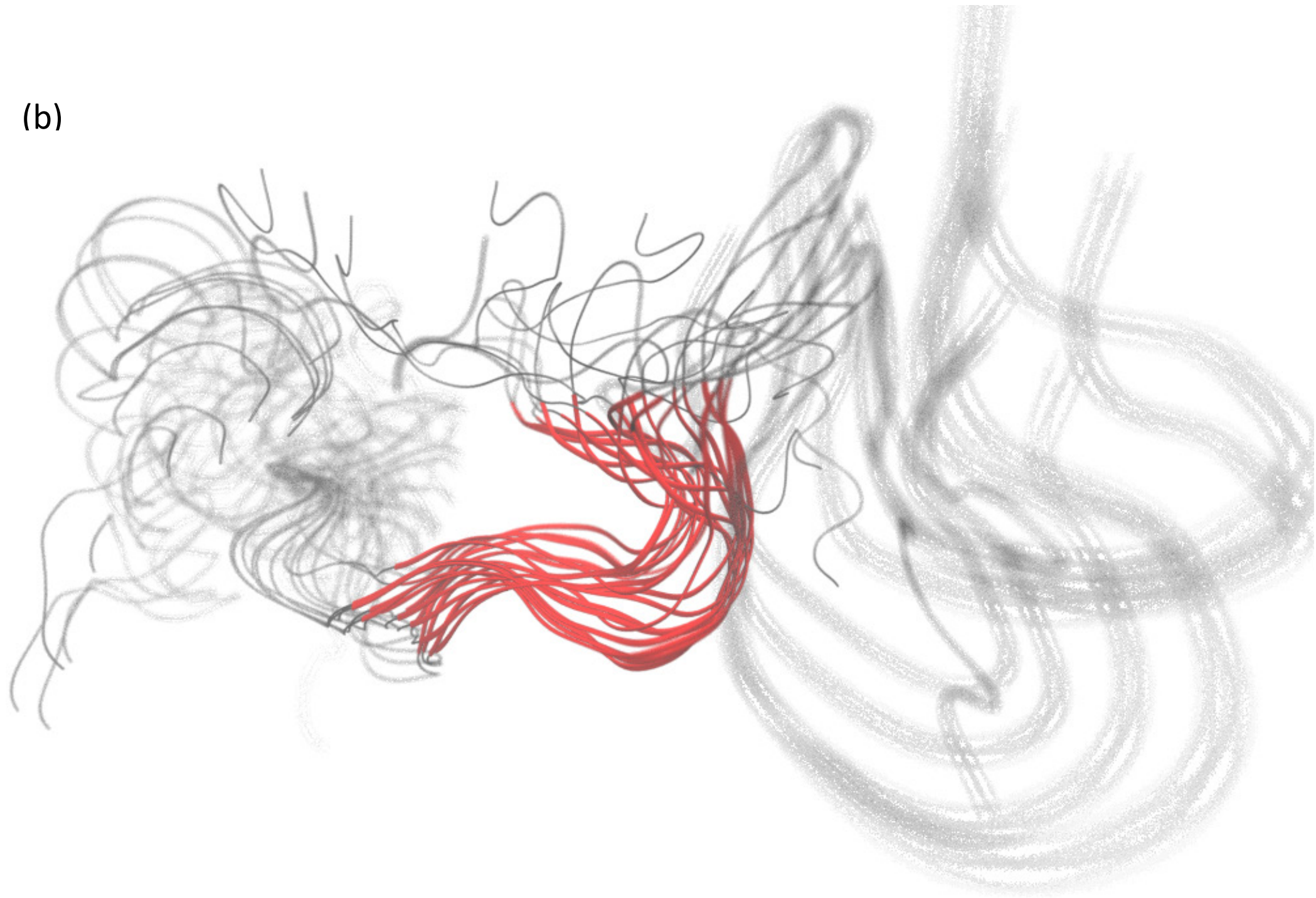

Figure S6: The time evolution of the secondary structure profile in the extended simulation at 310 K showed four distinct regions of persistent $\beta$-sheet like conformations, (a) Leu4-Gly5$_{200\text{-}201}$ - Leu4-Gly5$_{205\text{-}206}$ (highlighted by the blue ribbons) and (b) Ile4-Gly5$_{210\text{-}211}$ - Ile4-Gly5$_{214\text{-}215}$ (red ribbons) with reduced conformational flexibility. The trajectory was captured by overlaying multiple frames at 10 ns intervals.

Tang et al. (2016) Figure S7

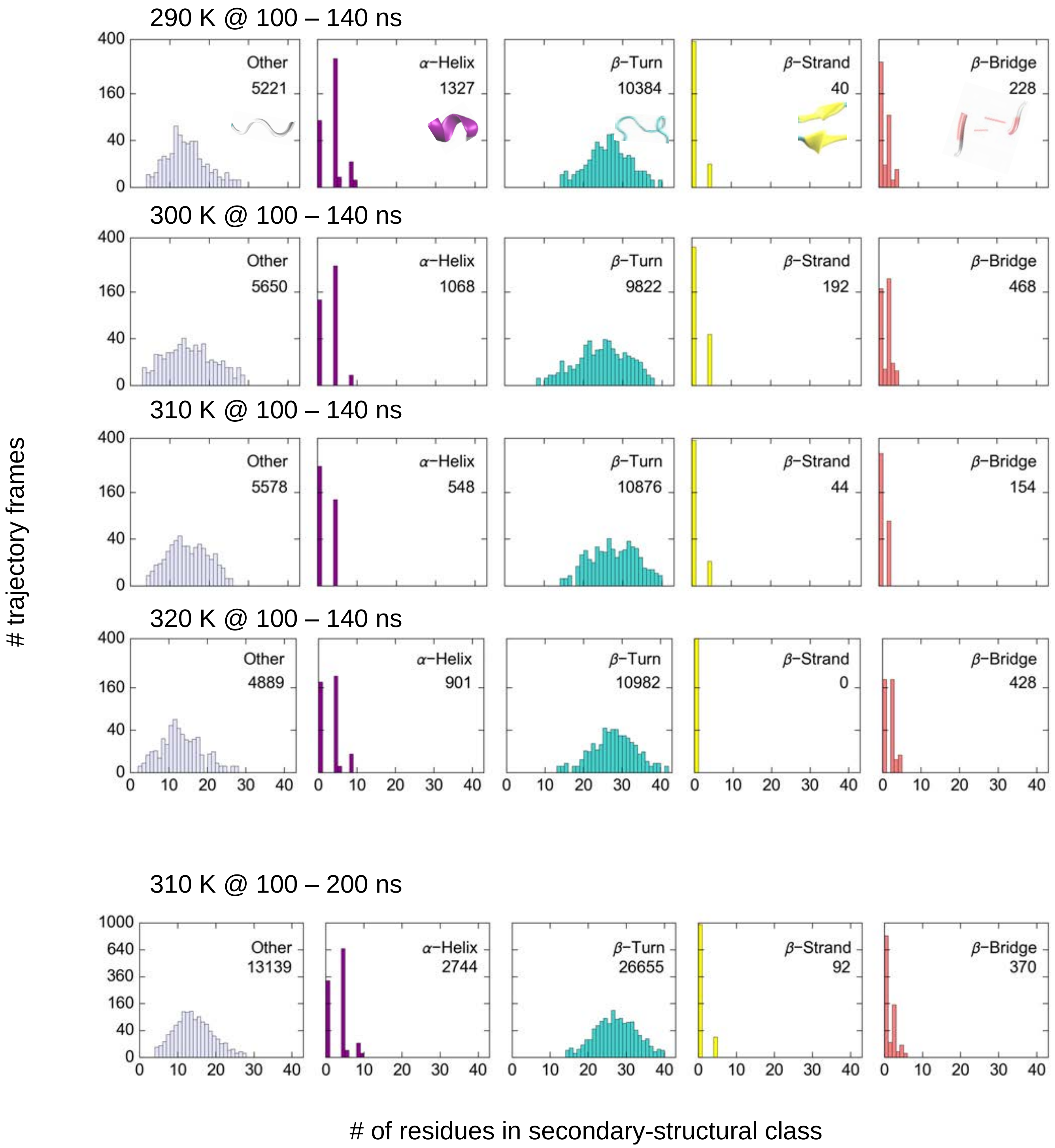

Figure S7: Secondary structural content of the ELP region across the 290–320 K temperature series for extended simulations (100 – 140 ns for 290 K, 300 K, and 320 K and 100 – 200 ns for 310 K) The conformational behavior of a single ELP monomer as a function of temperature. Numbers within the plot represent the total number of times that the secondary structure was observed in the simulation. Secondary structure cartoon representations in the thumbnails displayed in the first row match the colors in the histogram.

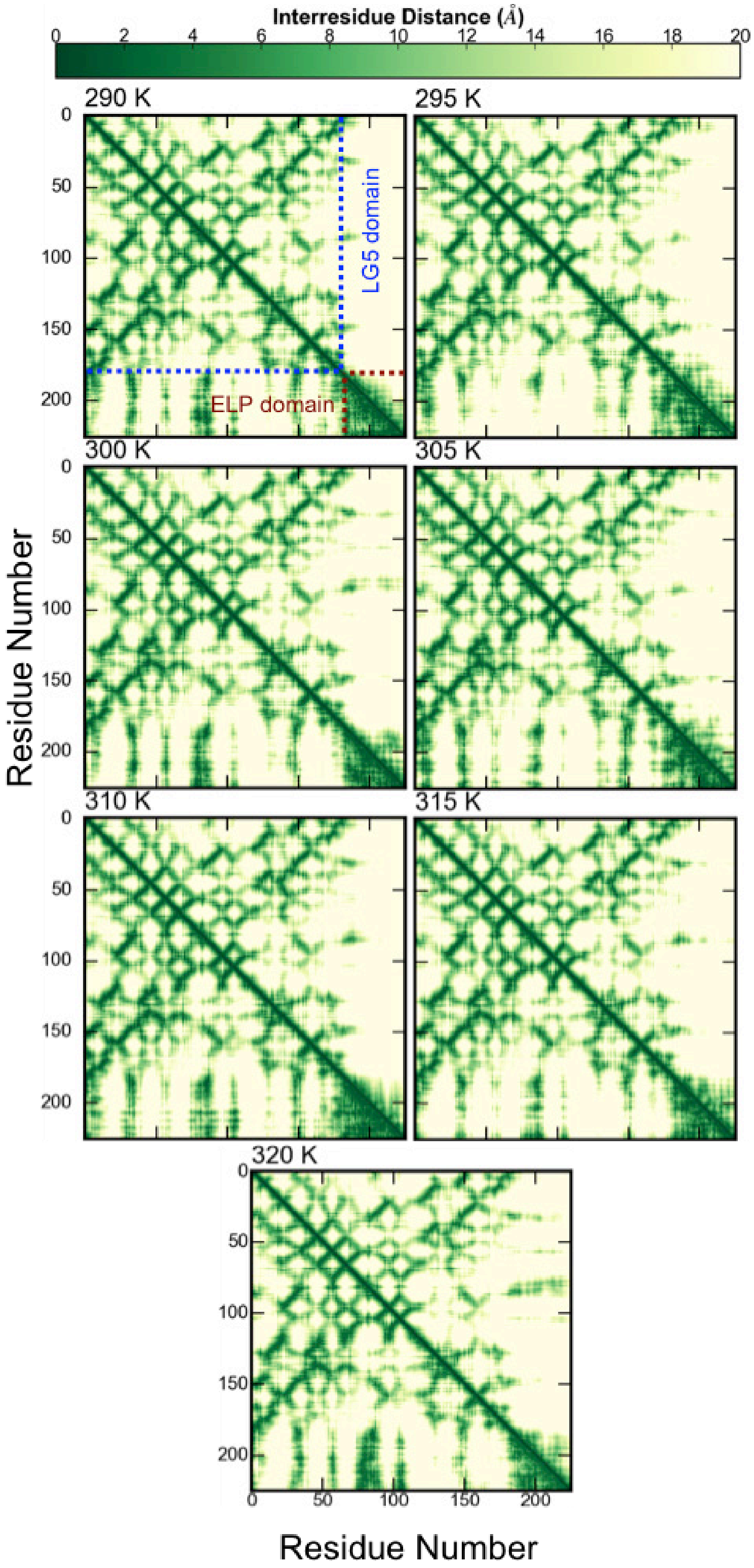

Figure S8: Protein contact maps of the dynamical interactions in the designed fusion suggest a lack of persistent LG···ELP interactions (for 290 K – 320 K).

Tang et al. (2016) Figure S9

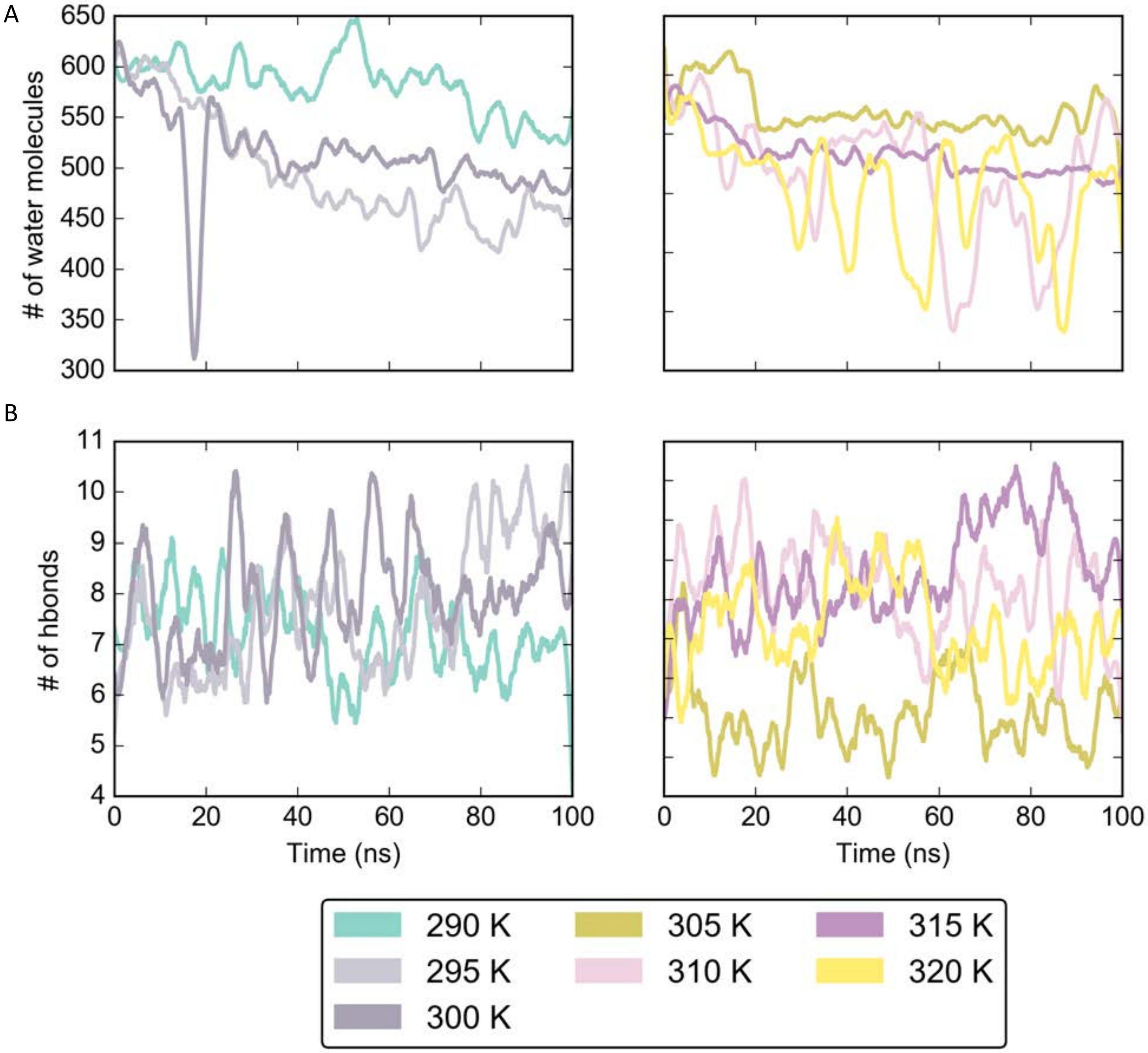

Figure S9 (A) Number of water molecules surrounding the ELP domain as a function of time. The abrupt drop in water molecules at 64 ns and 82 ns for 310 K corresponds to the formation of β-sheets. (B) Number of intramolecular hydrogen bonds as a function of time. All data was smoothed using a Savzky-Golay filter with a window size of 51 and 3rd order polynomial.

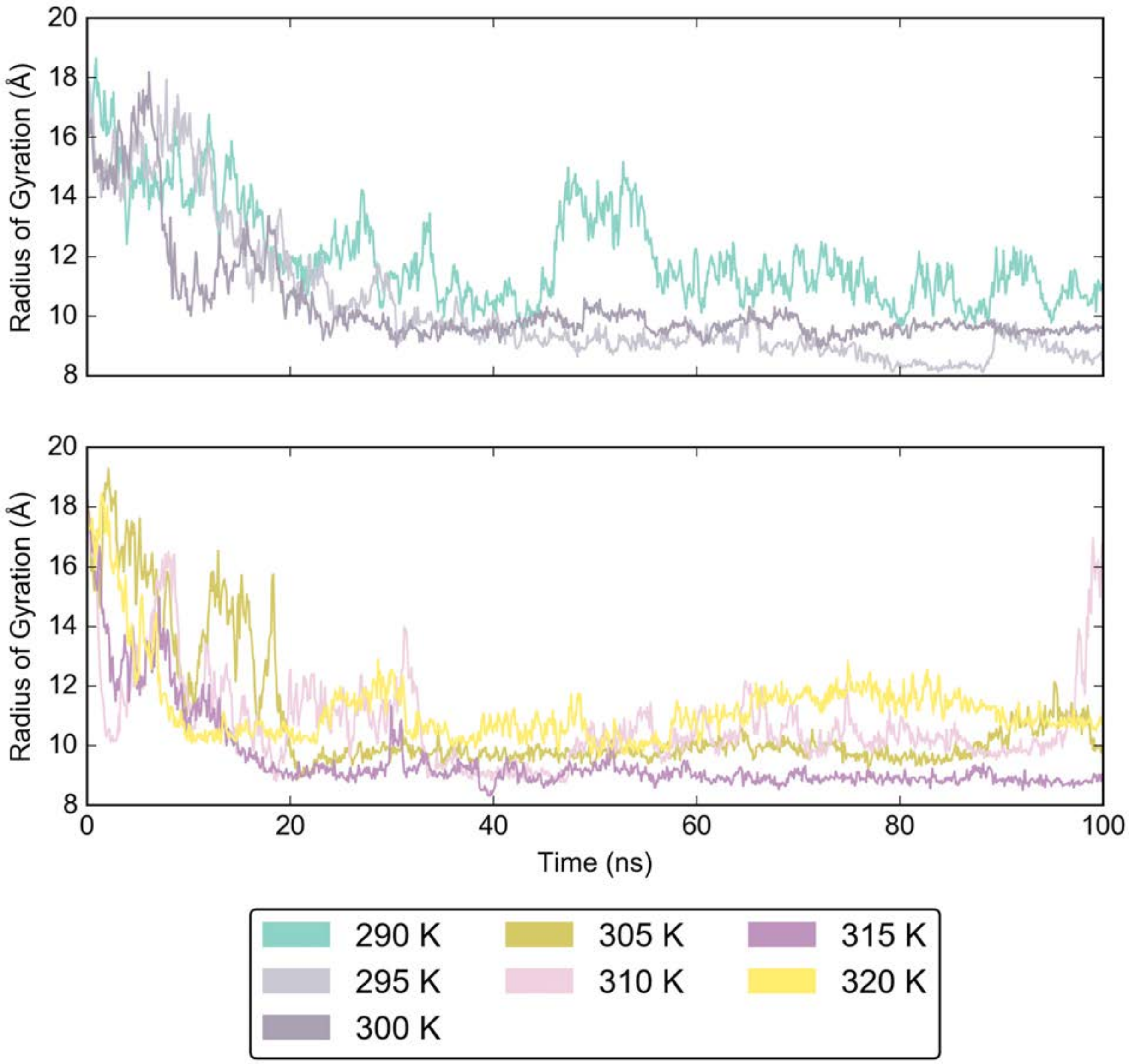

Figure S10: Time evolution of the radius of gyration simulated at different temperatures.

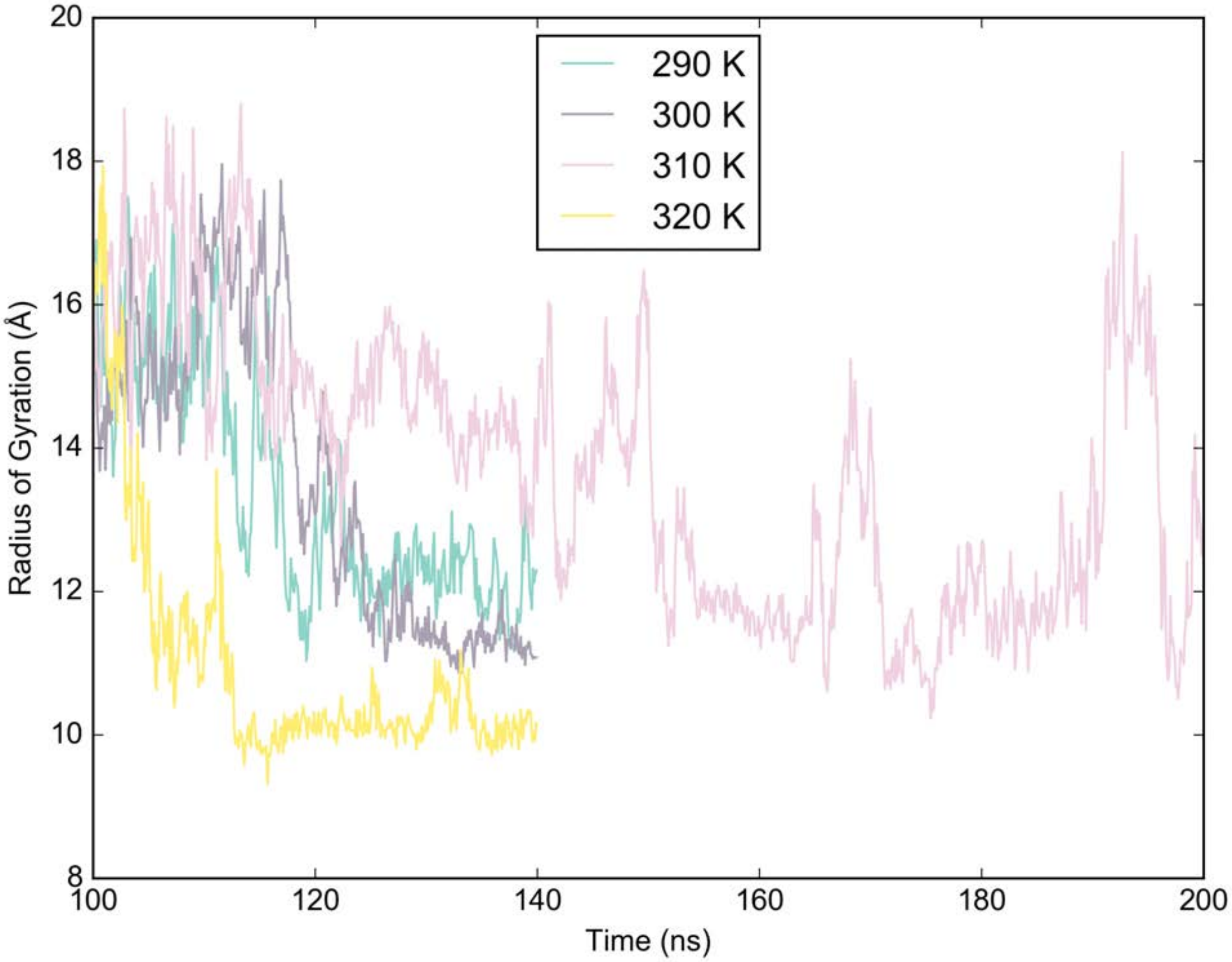

Figure S11: Time evolution of the radius of gyration at different temperatures for extended simulations (100 – 140 ns for 290 K, 300 K, and 320 K and 100 – 200 ns for 310 K)

Tang et al. (2016) Figure S12

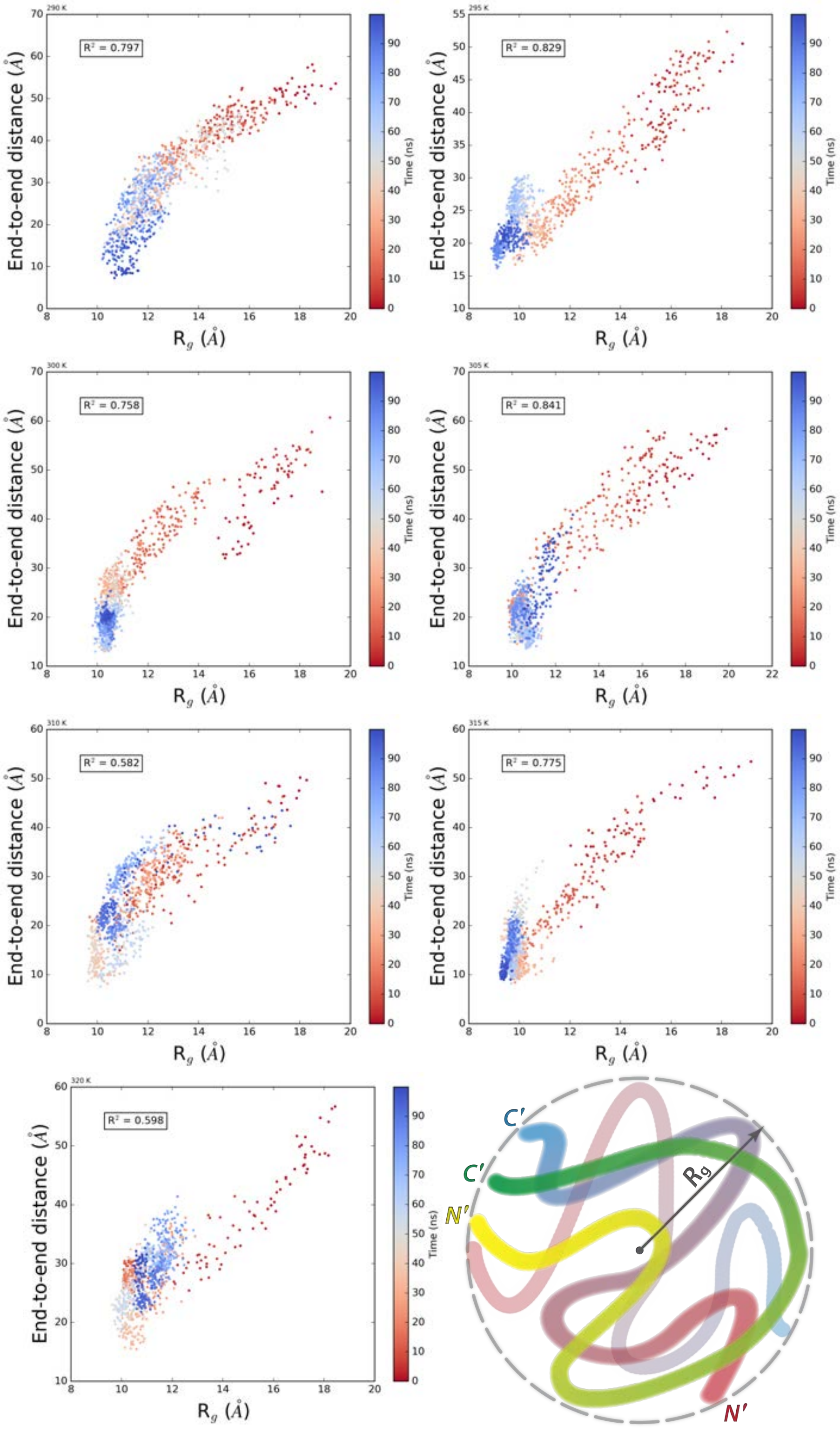

Figure S12: Correlation between the radius of gyration and end-to-end distance of the ELP domain. We computed a least-squares regression using SciPy's stats function. $R^2$ is the coefficient of determination. Colors correspond to the time steps in the simulation (red indicates first time step, blue is the last time step of the MD simulation). The following schematic represents the differences between the end-to-end distance (Euclidean distance between N'- and C'- termini) and radius of gyration ($R_g$) of a random coil.

Tang et al. (2016) Video Files

Video S1: Molecular dynamics simulation of the LG-ELP fusion protein at 310 K from 0 ns to 100 ns. The native conformation of the LG domain is maintained throughout simulation. α-helices colored purple, $3_{10}$ helices blue, *β*-strands yellow, the *β*-turn motif cyan and irregular coil regions white.

Video S2: Extended simulation of the LG-ELP fusion protein (from the final frame of the 100 ns run at 310 K) for 40 ns at 290 K showing a rapid decrease of β-sheet after 5 ns and a transition into a collapsed state. α-helices colored purple, $3_{10}$ helices blue, *β*-strands yellow, the *β*-turn motif cyan and irregular coil regions white.

Video S3: Extended simulation of the LG-ELP fusion protein (from the final frame of the 100 ns run at 310 K) for 40 ns at 300 K showing persistent *β*-sheet characteristics. α-helices colored purple, $3_{10}$ helices blue, *β*-strands yellow, the *β*-turn motif cyan and irregular coil regions white.

Video S4: Extended simulation of the LG-ELP fusion protein (from the final frame of the 100 ns run at 310 K) for 100 ns at 310 K exhibiting the same structural rearrangement behavior as the initial 100 ns run. α-helices colored purple, $3_{10}$ helices blue, *β*-strands yellow, the *β*-turn motif cyan and irregular coil regions white.